\documentclass[10pt,conference]{IEEEtran}
\IEEEoverridecommandlockouts
\usepackage{changepage}
\usepackage{tabularx}
\usepackage{subcaption}
\usepackage{multirow}
\usepackage{makecell}
\usepackage{tabularx}
\usepackage{orcidlink}

\usepackage[acronym,nohypertypes={acronym}]{glossaries}
\usepackage[table]{xcolor}
\usepackage{framed}
\usepackage{balance}
\makeatletter
 {\par\unskip\endMakeFramed}
\makeatother

\definecolor{formalshade}{rgb}{0.93,0.93,0.93}
\definecolor{darkblue}{rgb}{0.2, 0.2, 0.2}

\newenvironment{formal}{%
  \def\FrameCommand{%
    \hspace{1pt}%
    {\color{darkblue}\vrule width 2pt}%
    {\color{formalshade}\vrule width 4pt}%
    \colorbox{formalshade}%
  }%
  \MakeFramed{\advance\hsize-\width\FrameRestore}%
  \noindent\hspace{-1pt}%
  \begin{adjustwidth}{}{7pt}%
  \vspace{2pt}\vspace{2pt}%
}
{%
  \vspace{3pt}\end{adjustwidth}\endMakeFramed%
}

\newcounter{resultcounter}
\newenvironment{result}{\begin{formal}
    \refstepcounter{resultcounter}
    \noindent \hspace{-4pt}
}{
\end{formal}
}

\usepackage{tcolorbox}
\usepackage{changepage}
\newtcolorbox{textbox}[1]{
    sharp corners,
    boxsep=0mm,
    toptitle=2mm,
    lefttitle=0mm,
    colframe=black!3,
    colback=black!3,
    title={\rule[-2pt]{4.5pt}{10pt}\hspace*{1.5mm}#1},
    fonttitle=\bfseries\itshape\sffamily,
    coltitle=black,
    halign=flush left,
}

\usepackage{hyperref,url}
\usepackage{xspace}
\usepackage{lipsum}
\usepackage{cleveref}
\usepackage{tikz}
\usepackage{amssymb}
\newcommand{\etal}[1]{#1~\textit{et al.}\xspace}
\newcommand\ie{\emph{i.e.},\xspace}
\newcommand\eg{\emph{e.g.},\xspace}

\newcommand{\mm}[1]{{\small{\textsf{#1}}}}

\newcommand{\citefig}[1]{Figure~\ref{fig:#1}}

\newcommand\etc{etc.\xspace}

\newcommand{\citetable}[1]{Table~\ref{table:#1}}

\usepackage{pifont}%
\newcommand{\cmark}{\ding{51}}%
\newcommand{\xmark}{\ding{55}}%

\newcommand\hf{Hugging Face\xspace}
\newcommand{\tech}[1]{\mm{#1}}

\usepackage{booktabs}

\usepackage{flushend}
\newacronym{spl}{SPL}{{\emph{Software Product Line}}}
\newacronym{ttft}{TTFT}{\emph{Time to First Token}}
\newacronym{itl}{ITL}{\emph{Inter-Token Latency}}
\newacronym{vllm}{vLLM}{\emph{virtual Large Language Model}}
\newacronym{tgi}{TGI}{\emph{Text Generation Inference}}
\newacronym{moe}{MOE}{\emph{Mixture of Experts}}
\newacronym[
    \glslongpluralkey={\emph{Large Language Models}}, 
    \glsshortpluralkey={LLMs}
]{llm}{LLM}{\emph{Large Language Model}}

\definecolor{efflarge}{HTML}{FFCC99}
\definecolor{effmedium}{HTML}{FFE699}
\definecolor{effsmall}{HTML}{FFF2CC}
\definecolor{effnegligible}{HTML}{FFFFFF}
\newcommand{\marktext}[2]{\colorbox{#1}{\strut #2}}

\usepackage{duckuments}

\def\BibTeX{{\rm B\kern-.05em{\sc i\kern-.025em b}\kern-.08em
    T\kern-.1667em\lower.7ex\hbox{E}\kern-.125emX}}
\begin{document}

\title{Attention to Detail: Evaluating Energy, Performance, and Accuracy Trade-offs Across vLLM Configurations}

\author{
    \IEEEauthorblockN{%
        \emph{Nada}~\textsc{Zine}\orcidlink{0009-0005-4689-3697}$^*$,\,
        \emph{Clément}~\textsc{Quinton}\orcidlink{0000-0003-3203-6107},\,
        \emph{Romain}~\textsc{Rouvoy}\orcidlink{0000-0003-1771-8791}%
    }
    \IEEEauthorblockA{
        Univ. Lille, CNRS, Inria\\
    }

    \and

    \IEEEauthorblockN{%
        \emph{Tristan}~\textsc{Coignion}\orcidlink{0009-0003-2525-3637}$^*$%
    }
    \IEEEauthorblockA{
        Univ. Bordeaux, CNRS, LaBRI\\
    }

    \and

    \IEEEauthorblockN{%
        \emph{Vincenzo}~\textsc{Stoico}\orcidlink{0000-0002-3681-372X},\,
        \emph{Ivano}~\textsc{Malavolta}\orcidlink{0000-0001-5773-8346},\,
        \emph{Patricia}~\textsc{Lago}\orcidlink{0000-0002-2234-0845}%
    }
    \IEEEauthorblockA{
        Vrije Universiteit Amsterdam\\
    }

    \thanks{$^*$Corresponding authors: {\tt nada.zine@inria.fr}, {\tt tristan@coignion.fr}}
}

\author{
   \IEEEauthorblockN{\emph{Nada}~\textsc{Zine}\orcidlink{0009-0005-4689-3697}$^*$}
   \IEEEauthorblockA{
       Univ. Lille, CNRS, Inria\\
   }

   \and

   \IEEEauthorblockN{\emph{Tristan}~\textsc{Coignion}\orcidlink{0009-0003-2525-3637}$^*$}
   \IEEEauthorblockA{
       Univ. Bordeaux, CNRS, LaBRI\\
   }

   \and

   \IEEEauthorblockN{\emph{Vincenzo}~\textsc{Stoico}\orcidlink{0000-0002-3681-372X}}
   \IEEEauthorblockA{
       Vrije Universiteit Amsterdam\\
   }

   \and

   \IEEEauthorblockN{\emph{Ivano}~\textsc{Malavolta}\orcidlink{0000-0001-5773-8346}}
   \IEEEauthorblockA{
       Vrije Universiteit Amsterdam\\
   }

   \and

   \IEEEauthorblockN{\emph{Clément}~\textsc{Quinton}\orcidlink{0000-0003-3203-6107}}
   \IEEEauthorblockA{
       Univ. Lille, CNRS, Inria, IUF\\
   }

   \and
   
    \IEEEauthorblockN{\emph{Romain}~\textsc{Rouvoy}\orcidlink{0000-0003-1771-8791}}
   \IEEEauthorblockA{
       Univ. Lille, CNRS, Inria\\
   }

   \and

   \IEEEauthorblockN{\emph{Patricia}~\textsc{Lago}\orcidlink{0000-0002-2234-0845}}
   \IEEEauthorblockA{
       Vrije Universiteit Amsterdam\\
   }

   \thanks{$^*$Corresponding authors: {\tt nada.zine@inria.fr}, {\tt tristan@coignion.fr}}
}

\maketitle

\begin{abstract}
\glspl{llm} are reshaping how software is developed and maintained. 
They are typically deployed in production using inference engines such as vLLM, which can efficiently serve pre-trained, highly configurable models. 
While prior work has focused on model architectures and hardware acceleration, the impact of inference engine configuration on energy consumption, performance, and output quality remains poorly understood.

In this paper, we present a large-scale controlled study of three selected vLLM configuration options: attention kernel type, prefix caching, and chunked prefill.  
We evaluate all combinations of these configurations across 5 open-weight LLMs and 5 diverse inference tasks, totaling $9,000$ runs and $93,600$ measures. 
We analyze energy consumption, latency, and accuracy, and examine both main effects and interaction effects between configuration options and tasks.

Our results show that the studied configuration options significantly impact energy and performance, mainly driven by attention type and prefix caching, while chunked prefill has a limited effect under the default vLLM serving configuration and evaluated workloads.
These effects are highly model- and workload-dependent, and no configuration is universally optimal. 
We further show that model choice dominates global trade-offs, while configuration tuning provides  local improvements along the Pareto frontier. Unexpectedly, inference options can also affect model accuracy.
\end{abstract}

\maketitle
\glsresetall
\section{Introduction}
\label{sec:introduction}

\glspl{llm}, like GPT and Gemini, are among the technologies we can reasonably say have changed the world~\cite{eloundou2024gpts}.
They have rapidly become key enablers of innovation across natural language processing, creative content generation and virtual assistants, with broad applications in industry and education~\cite{bharathi2024analysis,murtaza2025impact}.

This growth raises concerns about their \textit{environmental impact}, due to their significant energy requirements, both in terms of embodied~\cite{vries-gao_recalibrating_2026} and operational emissions~\cite{stojkovic2024towards}.
Both training and inference phases are highly resource-intensive~\cite{jegham2025hungry}. \textit{Inference}, in particular, is attracting increasing attention, as \glspl{llm} receive millions of daily requests, accounting for more than $90$\% of the total computational workload associated with this type of machine learning models~\cite{fu2025llmco2,patel2024characterizing}.
To address these concerns, researchers have begun investigating various inference optimization strategies, either by proposing new methods to make \glspl{llm} inference more energy-efficient or by empirically evaluating existing approaches in terms of energy consumption, performance, and accuracy~\cite{ding2024sustainable}.

An \textbf{\glspl{llm} inference engine} is a specialized software system that efficiently execute pre‑trained \glspl{llm}, managing tokenization, memory allocation (\eg KV‑cache), scheduling, batching, \etc to deliver low‑latency and high‑throughput outputs at scale~\cite{park2025survey}.
Various inference engines for \glspl{llm} exist~\cite{llms_landscape}, including cloud-oriented frameworks such as \gls{vllm}~\cite{kwon2023efficient} and Hugging Face Transformers~\cite{wolf2019huggingface}, as well as local deployment solutions like Ollama~\cite{ollama} and llama.cpp~\cite{llamacpp}.
Among these, \gls{vllm} is the most widely used state-of-the-art \gls{llm} inference and serving engine~\cite{su2025seesaw}.
A key challenge in deploying \gls{llm} inference engines is their \textbf{large and complex configuration space}~\cite{park2025survey}, with tens of options, including temperature, top\_p, top\_k, \etc
These configuration choices can significantly affect runtime properties such as latency, throughput, energy consumption, and even accuracy.
However, properly configuring an \gls{llm} inference engine is not trivial since configuration options (i) might conflict with each other (\eg enabling KV‑cache and very long attention windows), (ii) are highly project- and context-dependent (\eg conversational AI agents require different decoding strategies than code generation
), and (iii) can lead to drastically different runtime behaviours in terms of, \eg energy consumption and accuracy~\cite{stojkovic2024towards}.

Despite active work on \gls{llm} inference optimization~\cite{solovyeva2026towards}, we still lack a systematic understanding of how inference-engine configuration choices affect energy, performance, and accuracy in realistic scenarios. 
In particular, existing work either focuses on isolated optimizations or evaluates configurations in limited settings, leaving unclear (i) how configuration options interact, (ii) how their effects vary across task types, and (iii) whether they impact model outputs~\cite{zine2026pimp}.

To address this gap, we conduct a \textit{controlled experiment}~\cite{wohlin2012experimentation} on the impact of three representative system-level vLLM configuration options on energy consumption, performance, and model accuracy.
We evaluate five highly downloaded open-weight \glspl{llm} from the \hf platform ranging from 3B to 32\,B, 
on $2,538$ tasks from five diverse datasets.

Because vLLM exposes tens of configuration options, exhaustively evaluating the entire configuration space is impractical. 
We therefore focus on exploring three widely used system-level options that target complementary stages of inference and allow a tractable full-factorial study across models and tasks:
(i) 3 types of \textbf{attention kernel} (\ie ~\tech{FlashAttention-2}, \tech{FlashAttention-3}, and \tech{FlashInfer}), which affect the low-level attention computation; (ii) \textbf{prefix caching} (on/off), which controls KV-cache reuse across shared prompt prefixes; and (iii) \textbf{chunked prefill} (on/off), which changes how long prompts are processed during the prefill phase. Each configuration is repeated $30$ times to capture metric fluctuations, leading to $9,000$ experimental runs.
All runs are executed in random order on $4$ nodes in a dedicated cluster. We collect vLLM runtime metrics and focus on (i) CPU and GPU energy usage, (ii) two complementary performance metrics (\ie time to first token and request inference time), and (iii) two accuracy metrics (\ie pass@k and average per-item score).
The experiment yields $93,600$ collected data points, which are then statistically analyzed. 

Our \textbf{results} show that \gls{vllm} configuration choices influence energy consumption, request latency, and Time to First Token (TTFT), although the magnitude of these effects depends strongly on the model and workload. Under the selected \gls{vllm} serving configuration and evaluated workloads, attention type and prefix caching emerge as the most impactful \gls{vllm}-specific options, while chunked prefill has a limited effect.

No single setup works best for all models or tasks, and interaction effects tend to be highly context‑specific.
Moreover, task types are the primary determinant of energy consumption and performance; attention type and prefix caching exhibit significant interactions with task types. 
Attention type varies substantially across task types, often introducing energy–latency trade-offs.
The choice of \gls{llm} dominates the energy-latency-accuracy Pareto front, while no single \gls{vllm} configuration emerges as universally optimal across task types.

The \textbf{main contributions} of this paper are:

(i) we go beyond the state of the art by providing empirical evidence that the selected vLLM options affect inference energy usage and performance, with task-dependent effects;
(ii) we find that configuration options such as attention type and prefix caching interact, meaning that the impact of one cannot be assessed in isolation; for example, the latency gain from prefix caching depends on the attention backend in use;
(iii) we show that model choice dominates global trade-offs, while configuration options provide only local optimizations along the Pareto frontier;
(iv) we reveal that inference configurations can unexpectedly affect model accuracy, challenging the assumption that such parameters are purely system-level optimizations;
(v) we release a full replication package~\cite{replicationpackage} to support reproducibility and future research.

The \textbf{target audience} includes:
(i) \textit{developers} integrating vLLM-served \glspl{llm} in their software systems with energy, performance, and accuracy in mind,
(ii) \textit{\gls{llm} inference engine vendors} to better understand how configuration options interact and improve their inference engines,
and (iii) \textit{software engineering researchers} seeking inspiration for their future research based on the results and implications of our study.
\section{Related Work}
\label{sec:related-work}

The energy consumption of \glspl{llm} inference has recently gained significant attention, as inference is increasingly considered as the major contributor to the overall footprint of deployed AI systems~\cite{wu2022sustainable}.
Prior work has shown that inference efficiency depends on multiple factors from different levels.

A first set of studies focuses on hardware and infrastructure. \etal{Patel} compared inference serving at the server and cluster levels~\cite{patel2024characterizing}. \etal{Stojkovic} studied the impact of task type, batching, and model parallelism in \gls{vllm}-based inference environments~\cite{stojkovic2024towards}. In later work, they analyzed the effects of varying the number of \gls{llm} instances, model parallelism, and GPU frequency settings, and proposed an energy-management framework for inference environments \cite{stojkovic2025dynamollm}. \etal{Lazuka} evaluated \glspl{llm} inference across multiple GPUs and proposed a predictive model to recommend the most cost-effective hardware for unseen \glspl{llm}~\cite{lazuka2024llm}.

A second set of studies focuses on models and tasks. \etal{Samsi} studied the impact of GPU settings, LLaMA model size, input data, batch size, and generation length~\cite{samsi2023words}. \etal{Luccioni} compared architectures, model sizes, and task types~\cite{luccioni2024power}. Other work examined the effects of DVFS, batch size, model choice, workload characteristics, prompt and output length, quantization, and serving configuration on energy, latency, and quality~\cite{maliakel2025investigating,wilkins2024offline,husom2025sustainable,delavande2026understanding,kolovskaSmallPromptsBig2025,fu2025llmco2}.

A third set focuses on decoding-level factors. Several studies examined the effect of decoding hyperparameters on GPU energy consumption, generation quality, and performance across models, quality metrics, and tasks~\cite{shi2402thorough,arias2025decoding,nik2025impact}. \etal{Zine} proposed variability models to evaluate and predict the energy consumption, performance, and output quality associated with Hugging Face Transformers generation hyperparameters~\cite{zine2026pimp}.

Finally, a fourth set focuses on inference serving factors. Prior work studied the effects of concurrency, request patterns, quantization, number of GPUs, inference hyperparameters, decoding strategies, GPU architectures, online versus offline serving, and model parallelism across systems such as \gls{tgi}, \gls{vllm} and Hugging Face Transformers~\cite{coignion2024green,martinez2025impact,fernandez2025energy,pronkBenchmarkingEnergyEfficiency2025,niuEnergyEfficientExhaustive2025}.

Overall, prior work shows that \gls{llm} inference efficiency depends on many factors. However, these factors are often studied in isolation or within a limited set of interactions, leaving their combined effects across models and task types heavily understudied.
We fill this gap by studying the main effect and the interactions between three system-level \gls{vllm} options targeting different parts of the inference pipeline: attention computation, cache reuse, and prefill scheduling.
\section{Study Design}\label{sec:methods}
We followed established empirical software engineering and energy measurement guidelines~\cite{wohlin2012experimentation,shull2007guide,FGCS_2024,jagannadharaoBeginnersGuidePower2024}. We also provide a complete \textit{replication package} \cite{replicationpackage} with the experimental pipeline, intermediate artifacts, collected measurements, and analysis scripts.

\subsection{Goal \& Research Questions}\label{sec:goal}
We define the goal of the study using the Goal-Question-Metric (GQM) framework by \etal{Basili}~\cite{basili1988tame}. The \textbf{goal} of this study is to analyse \textit{LLM inference engine configurations}
for the purpose of \textit{understanding}
their impact on \textit{energy consumption, performance, and accuracy}
from the point of view of \textit{software developers, LLM inference engine vendors, and Software Engineering researchers}
in \textit{vLLM deployments}.

To achieve it, we evaluate \gls{vllm} across three configurations using $5$ \glspl{llm} and $5$ datasets that represent different task types.
For each experimental trial, we measure the energy consumption and performance (across diverse metrics) of the vLLM instance, as well as the accuracy of the LLMs' outputs.
The research questions of this study are as follows.

\noindent $\mathbf{RQ_1}$ -- For a given \gls{llm} and task type, how do \gls{vllm} configuration options and their interactions affect energy consumption, performance, and LLM's accuracy?
This RQ assesses whether the studied \gls{vllm} options exhibit different behaviors in terms of energy, performance (total latency, \gls{ttft}), and LLM's accuracy.

\noindent $\mathbf{RQ_2}$ -- What interaction effects exist between \gls{vllm} configurations and task types on energy consumption, performance, and accuracy?
In this RQ, we investigate whether different \gls{vllm} configurations interact with task characteristics in ways that affect energy consumption, latency, \gls{ttft}, and accuracy. 

\noindent $\mathbf{RQ_3}$ -- For a given type of task, what Pareto‑optimal \gls{vllm} configurations balance energy consumption, performance, and LLMs' accuracy?
This RQ focusses on how to best configure a \gls{vllm} instance in terms of resource usage (\ie energy and CPU/GPU usage), performance (total latency) and LLM's accuracy when having a fixed type of task and deploying varying \glspl{llm}. 

\subsection{Selection of vLLM Configuration Options}

As \gls{vllm} exposes tens of configuration options~\cite{vllm_docs}, exhaustively evaluating its configuration space across multiple models, tasks and repetitions would be impractical. We therefore focus on system-level options that satisfy three criteria: (i) they are exposed to practitioners and can be changed without modifying model weights, (ii) they target runtime mechanisms likely to affect energy consumption, latency or benchmark scores, and (iii) they are model-independent, \ie they do not directly influence the choice of the \gls{llm} being served by the vLLM instance.
Based on these criteria, we select three complementary options: (1) attention kernel, (2) prefix caching, and (3) chunked prefill.
Together, these options cover attention computation, cache reuse, and prefill scheduling, while keeping the design tractable for a full-factorial controlled experiment across models and tasks. 

\noindent \textbf{Attention kernel} refers to the low-level GPU implementation of the attention mechanism in transformer models. We consider three attention
kernels: \tech{FlashAttention-2}~\cite{dao2022flashattention},
\tech{FlashAttention-3}~\cite{dao2023flashattention}, and \tech{FlashInfer}~\cite{ye2025flashinfer}.
They focus on optimizing the attention mechanism by minimizing GPU memory accesses, thereby improving inference performance.

\noindent \textbf{Prefix caching} improves inference efficiency by caching the KV store of prompts during prefill  reducing processing time and costs.
It works best when prompts share a common prefix, allowing the system to quickly access cached results for repeated queries.
Prefix caching is expected to shorten the prefill phase and lower the \gls{ttft}~\cite{vllm_docs}. We evaluate this option with two settings, \tech{on} and \tech{off}.

\noindent \textbf{Chunked prefill} improves long-prompts processing by splitting the prefill phase into smaller chunks. Under the default \gls{vllm} setup used here, chunking is only activated when prompts exceed the default \texttt{max\_num\_batched\_tokens} threshold, set to $8192$. Consequently, we expect its impact mainly on long-context workloads. We evaluate this option with two settings, \tech{on} and \tech{off}.

\subsection{Selection of LLMs}
We selected \glspl{llm} from Hugging Face, the most popular platform for hosting ML-based projects~\cite{ait2025suitability}. We searched text-generation models by downloads in October 2025~\cite{huggingfaceTextGeneration}), then retained models that: (i) have at least $3$\,B parameters, excluding micro-LLMs; (ii) have fewer than $40$\,B active parameters to fit into a single GPU (see Section~\ref{sec:experiment_execution});
(iii) are open weights for replicability;
(iv) use dense architecture, excluding out-of-scope \gls{moe} models, which do not share the same attention kernels as dense models;
and (v) are instruction-tuned rather than task-specific.

We then iteratively selected the most downloaded remaining model family, kept its latest dense collection, and chose the largest model below $40$,B active parameters and, when available, the smallest above $3$,B. Selected families were removed from subsequent iterations until reaching three distinct families. This strategy favors recent models from popular open-weight families, while remaining feasible and replicable.

\begin{table}[t]
    \footnotesize
    \caption{LLMs considered in our study}
    \vspace{-2mm}
    \resizebox{1\columnwidth}{!}{
    \begin{tabular}{llrrr}
        \toprule
        \textbf{LLM}          & \textbf{Family} & \textbf{Size} & \textbf{Context size} & \textbf{Downloads} \\
        \midrule
        Qwen3-32B             & Qwen            & 32B           & 32,768                & 4,319,243          \\
        Qwen3-4B              & Qwen            & 4B            & 32,768                & 6,159,027          \\
        Magistral-Small-2509  & Mistral         & 24B           & 32,768                & 20,807             \\
        Llama-3.1-8B-Instruct & Llama           & 8B            & 131,072               & 7,360,488          \\
        Llama-3.2-3B-Instruct & Llama           & 3B            & 131,072               & 3,857,887          \\ \bottomrule
    \end{tabular}
    }
    \label{table:llm-description-table}
    \vspace{-5mm}
\end{table}
\autoref{table:llm-description-table} presents the final list of selected models.
For clarity, in the rest of the paper, we refer to each model by its family name and size only, \eg \tech{Qwen-32B} for \tech{Qwen/Qwen3-32B}.

\subsection{Selection of Task}
We select five datasets to cover complementary task characteristics relevant to the studied vLLM options, including prompt/output length, repeated prompt prefixes, single versus multi-turn interactions, and ground-truth availability. This selection allows us to detect interactions between task characteristics and vLLM configurations while keeping the scope of the study explicit.
An overview of the selected datasets is presented in \autoref{table:dataset-description}.

\noindent \textbf{AssistantTraces (AT)} contains GitHub Copilot generations from $20$ coding sessions with different developers~\cite{coignion2024green}.
We evaluate the first $10$ turns of each session in sequential batches.
The dataset reflects the typical tasks handled by a code-assistant server.

\noindent \textbf{EvoEval (EE)} contains $100$ programming problems from HumanEval with subtle changes in the requirements~\cite{xia2024top}.
For each problem, we request the \glspl{llm} to generate $5$ solutions. 
This dataset consists of cases where developers use \glspl{llm} to implement functions from predefined documentation and requirements.

\noindent \textbf{LongBench-v2 (LB)} evaluates long-context tasks requiring deep reasoning across real-world multitasks~\cite{bai2025longbench}.
Since most models have limited context windows, we use the \tech{short} subset ($<32k$ words), sample $30$ tasks, and format them with the benchmark's zero-shot template. LB captures workloads dominated by prefill, with limited decoding.

\noindent \textbf{Natural Questions (NQ)} contains users' questions issued to Google search with answers found on Wikipedia by annotators~\cite{kwiatkowski2019natural,lighteval}.
We sample $1,500$ questions from $7,842$ test questions.
To limit answer length and avoid right-tailing~\cite{hu2025taming}, we set \tech{max\_new\_tokens} to $512$.
Prompts use a concise QA format with a question and an \textit{Answer:} cue, tasking LLMs with a simple QA without external documents.

\noindent \textbf{WildChat (WC)} contains $650,000$ real user-ChatGPT conversations across diverse languages and prompts~\cite{zhao2024wildchat}.
We sampled $200$ multi-turn conversations, limit them to five turns, and organized them by turn number into five batches.
For subsequent turns, we used the dataset's original assistant responses as conversation history.
The output limit was set to $32,000$ to simulate a ChatGPT-like multi-turn setting.

We evaluate accuracy only on EE and LB, which provide ground-truth answers and established scoring procedures. AT and WC lack ground-truth, while NQ contains data used for training rather than for benchmarking.

\begin{table}[t]
    \footnotesize
    \caption{Overview of the task datasets used in this study.}
    \vspace{-2mm}
    \resizebox{1\columnwidth}{!}{
    \begin{tabular}{lrrrrr}
        \toprule
                                       & \textbf{AT} & \textbf{EE} & \textbf{LB} & \textbf{NQ} & \textbf{WC} \\ \midrule
        Original \#tasks               & 20          & 100         & 503         & 7,842       & 650,000     \\
        \#Tasks sampled                & 20          & 100         & 30          & 1,500       & 200         \\
        Avg. Prompt Length             & \makecell[r]{475 \\ $\pm$8.5}   & \makecell[r]{155 \\ $\pm$2}    & \makecell[r]{25,583 \\ $\pm$1,734} & \makecell[r]{35.8 \\ $\pm$.42} & \makecell[r]{818 \\ $\pm$4.6}  \\
        Avg. Output Length             & \makecell[r]{1,096 \\ $\pm$631} & \makecell[r]{159 \\ $\pm$47}   & \makecell[r]{435 \\ $\pm$214}      & \makecell[r]{172 \\ $\pm$123}  & \makecell[r]{1,060 \\ $\pm$521} \\
        Repeated prompt prefixes       & \cmark      & \cmark      & \xmark      & \xmark      & \cmark      \\
        Multi-turn                     & \cmark      & \xmark      & \xmark      & \xmark      & \cmark      \\
        Can be scored?                 & \xmark      & \cmark      & \cmark      & \xmark      & \xmark      \\
        \bottomrule
    \end{tabular}
    }
    \label{table:dataset-description}
    \vspace{-4mm}
\end{table}

\subsection{Experimental Variables}\label{sec:variables}
The independent variables of this study are
(i) the \textbf{attention kernel} (\ie \tech{FlashAttention-2}, \tech{FlashAttention-3}, and \tech{FlashInfer}),
(ii) the \textbf{prefix caching} (on/off),
(iii) the \textbf{chunked prefill} (on/off),
(iv) the \textbf{LLM} (see \autoref{table:llm-description-table}),
(v) the \textbf{task type} for the LLM (see \autoref{table:dataset-description}).
The \textbf{used LLM} (see \autoref{table:llm-description-table}) is the \underline{blocking factor} in our experiment.
The data collected during this experiment are grouped by different independent variables (and the blocking factor), depending on the specific RQ being answered (see~\autoref{sec:goal}). The dependent variables represent the specific quality focus being considered---\ie energy, performance, or accuracy. We also collect internal \gls{vllm} runtime metrics to support interpretation; these are available in our replication package~\cite{replicationpackage}.

\noindent \textbf{Energy}.
During the inference process, we sample the power draw of the CPU and GPU via \texttt{perf} and \texttt{nvidia-smi}, respectively, both with a sampling rate of 1Hz.
Energy consumption ($E$, in Joules) is then computed by using the classical $E = W \times T$ formula, where $W$ is the mean power draw of the system in Watts and $T$ is the duration of the experimental run in seconds~\cite{FGCS_2024}. We use direct software-level sampling to control the measurement window, avoid tool-specific aggregation heuristics, and keep the procedure independent of any specific higher-level energy framework.

\noindent \textbf{Performance}.
We measure performance by collecting two metrics via \gls{vllm}: (i) \textit{\gls{ttft}}, defined as the delay between sending a request to the vLLM instance and receiving the first generated token, and (ii) \textit{Request Inference Time}, which represents the end-to-end duration of the inference request from submission to completion.

\noindent \textbf{Accuracy}.
For EvoEval, we assess the functional correctness of the generated code using the \textit{pass@k} metric~\cite{chen2021evaluating}, which estimates the probability that at least one of the $k$ generated samples passes all test cases.
For LongBench, accuracy is computed as the \textit{average per-item score}; specifically, a prediction receives a score of 1 if it exactly matches the ground-truth answer, 0 otherwise, with a compensation of 0.25 assigned when no extractable answer is produced~\cite{bai2025longbench}.

\subsection{Data Analysis}\label{sec:data_analysis}
Following \etal{Wohlin}'s guidelines~\cite{wohlin2012experimentation}, we analyse $RQ_1$ and $RQ_2$ through \emph{data exploration}, \emph{hypothesis testing}, and \emph{effect size estimation}, while $RQ_3$ uses a Pareto-frontier approach to identify non-dominated trade-offs~\cite{pareto}.

\noindent \textbf{Data exploration}.
We characterize the collected measures using descriptive statistics, tables and boxplots to compare visually the groups.

\noindent \textbf{Hypothesis testing}.
We assess the normality of the distributions of our groups by using density plots and the Shapiro-Wilk test ($\alpha$ = 0.05).
Since the normality assumption is not satisfied in most cases, we use the Aligned Rank Transform (ART) ANOVA~\cite{wobbrock2011aligned}, a non-parametric extension of factorial ANOVA.
$RQ_1$ and $RQ_2$ use the same dependent variables (\textit{energy consumption}, \textit{end-to-end latency}, and \gls{ttft}).
For $RQ_1$, for each model and task type, we conduct a three-factor ART ANOVA with \textit{attention type} ($3$ levels), \textit{prefix caching} ($2$ levels), and \textit{chunked prefill} ($2$ levels), studying both main effects and their interactions.
For $RQ_2$, for each model, we conduct three separate two-factor ART ANOVA analyses---one per configuration factor, each paired with \textit{task type} ($5$ levels)---to assess configuration×task-type interactions.
We adjust p-values with Holm-Bonferroni~\cite{holm1979simple} to reduce false positives across comparisons, while controlling the family-wise error rate less conservatively than other methods. We use ART contrasts for significant effects.

\noindent \textbf{Effect size estimation}.
We use Cliff's Delta~\cite{cliff1993dominance} for non-parametric effect size estimation and interpret its results as proposed by Vargha and Delaney~\cite{vargha2000critique}.

\noindent \textbf{Pareto frontier calculation}.
To compute the frontier, we first aggregate repeated runs of the same configuration into a single point, using the median for energy and latency metrics, and the mean for accuracy when available. We then compare configurations by minimizing energy and latency and maximizing accuracy. A configuration belongs to the Pareto frontier if it is not dominated by other configurations.

\section{Experiment Execution}\label{sec:experiment_execution}
When combining our selected configuration factors and the models described in  \citetable{llm-description-table}, we obtain $3 \times 2 \times 2 \times 5 = 60$ unique configurations.
Each of these configurations is evaluated on five task types presented in \citetable{dataset-description} and is executed $30$ times to account for measurement variability.
As such, our experiment is comprised of $60 \times 5 \times 30 = 9,000$ individual runs. In total, our experiments consumed approximately (1.44 $\times 10^6$)~kJ $\approx$400~kWh (enough to power a Western-European home for 28 days). %
Each individual run consists of loading the LLM for a given configuration, running a warmup and calibration phase, and then submitting the dataset's prompts to vLLM's offline inference interface. We use the offline interface to keep prompt submission, batching, and cache state controlled across repetitions. For single-turn datasets (EE, LB, NQ), all prompts are submitted as a single batch; vLLM internally schedules and processes them concurrently.
For multi-turn datasets (AT, WC), one batch per conversation turn is submitted sequentially, with each batch containing one prompt per active conversation.
When selecting the next configuration and repetition to run, the selection is random to avoid running all repetitions of a given configuration on the same day and node.
To mitigate cold-start effects, we included a warmup before each configuration evaluation.
The warmup consists of three short prompts (a few words each) and one long prompt (which maximizes context size), all run sequentially.
Evaluations were preceded by a 30-second calibration phase, during which we measured the machine's idle energy consumption.
Between dataset evaluations, the KV cache and prefix cache are explicitly reset to prevent state leakage across runs, ensuring that prefix-caching configurations start each run from a cold cache.
Energy is measured exclusively during the generation phase; model loading and warmup are excluded from the reported energy figures.
Across all experiments, idle consumption measurements exhibited a coefficient of variation of $0.024$, and repeated energy measurements of identical configurations had a coefficient of variation of $0.043$, indicating high measurement stability.

All experiments run on a cluster of $4$ identical
nodes with an AMD EPYC 7513 processor, 512\,GiB of memory, and
\(4\) Nvidia A100-SXM4-40GB GPUs.
The system ran Debian~11 with the Linux 5.10.0-28-amd64 kernel.
Lastly, we used vLLM version 0.10.2, \tech{FlashInfer} version 0.4.1, and cuda 12.2.128.
All configurations use the same inference parameters: (i) \tech{temperature} = 0.4, a commonly adopted value in LLM inference studies~\cite{chen2021evaluating}; (ii) \tech{top-p} = 0.95, following the nucleus sampling setting recommended by Holtzman \etal~\cite{holtzmanCuriousCaseNeural2020}; and (iii) \tech{max-tokens} = $2,048$, an upper bound on the output tokens that limits truncation while avoiding unusually long generations when models fail to stop. Unless stated otherwise, all remaining \gls{vllm} serving and decoding parameters were kept at their default values.
We intentionally adopted these defaults to remain representative of common vLLM deployments while isolating the effects of the studied configuration options.

\section{Results}\label{sec:results}
\subsection{Impact of \gls{vllm} Configurations (RQ1)}\label{subsec:rq1}

\subsubsection{\textbf{Data Exploration}}
\citefig{conf_comparison_box_plot} shows that energy varies across \gls{vllm} configurations for all models, although the importance of this effect strongly depends on the model.
\begin{figure}[t]
   \centering
   \includegraphics[width=\linewidth]{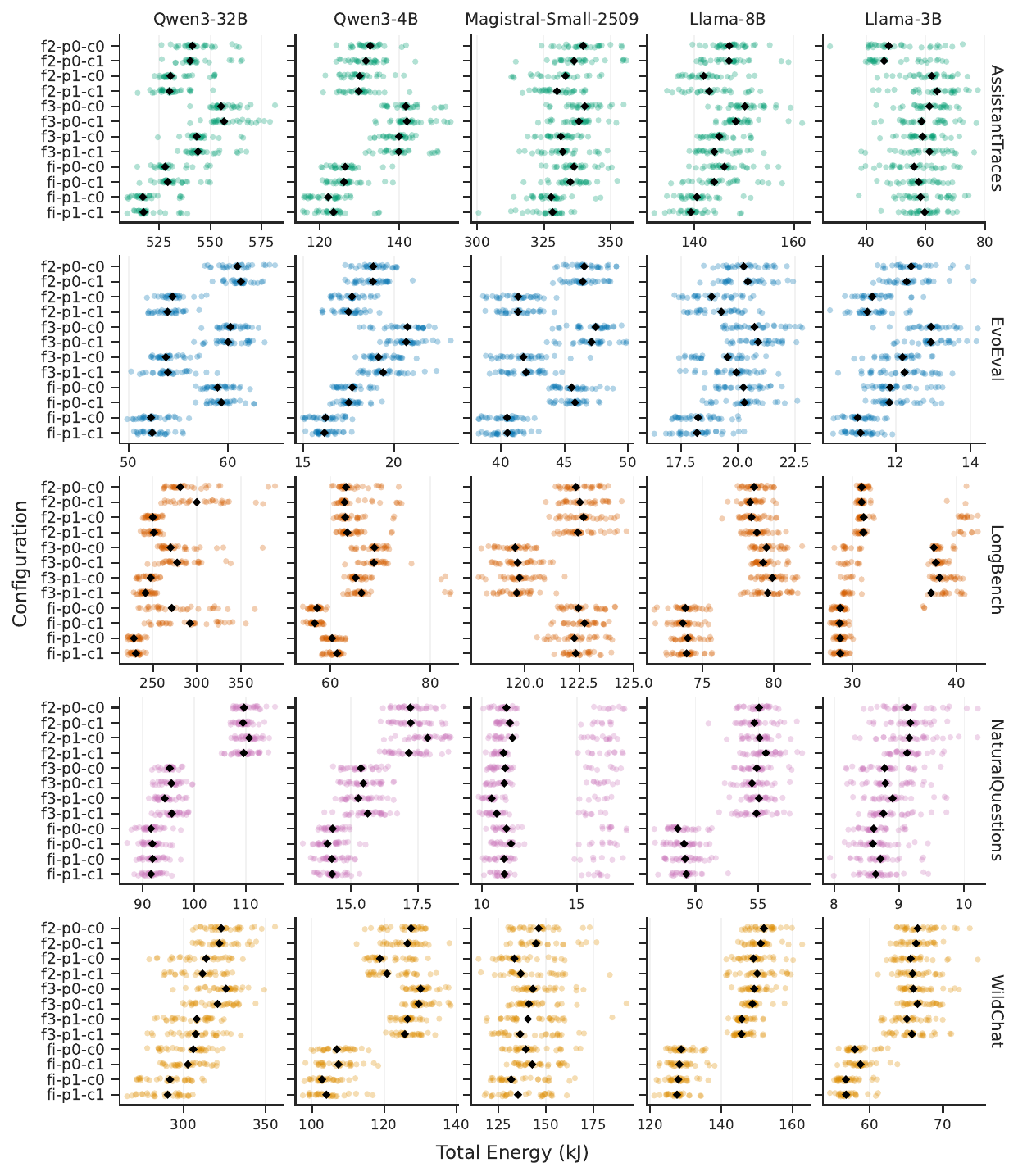}
   \vspace{-3mm}
    \caption{Total energy consumption (kJ) by model, task, and \gls{vllm} configuration. Points are runs; black dots show medians. IDs encode attention kernel (f2=\tech{FlashAttention-2}, f3=\tech{FlashAttention-3}, fi=\tech{FlashInfer}); p/c encode prefix caching/chunked prefill (0/1=off/on).}
   \vspace{-4mm}
   \label{fig:conf_comparison_box_plot}
\end{figure}

We observe that for certain model–task pairs, the resulting distributions differ notably across configurations, indicating that configuration choices can significantly impact energy consumption, whereas for others the distributions overlap much more, showing lower sensitivity.
On LB, for example, \tech{Qwen-32B} spans 220\,kJ to more than 388\,kJ per prompt, while \tech{Llama-3B} ranges from about 28 to 42\,kJ.
These examples illustrate two distinct effects: model size mainly shifts the energy baseline, whereas configuration sensitivity varies by model--task pair and does not simply scale with model size, as \tech{Qwen-32B}, despite its highest overall consumption, varies by up to 30\% across configurations compared with about 39\% for \tech{Llama-3B}.
A similar trend is observed for \gls{ttft} and total latency, whose sensitivity to configuration choices also depends on the model and task (the corresponding plots are provided in the replication package).
However, the effect's importance differs substantially across metrics: median differences reach 38.8\% for energy consumption, 80.6\% for request latency, and 324.0\% for \gls{ttft}, indicating a particularly strong effect on \gls{ttft}.

\begin{table}[t]
   \footnotesize
   \centering
   \setlength{\tabcolsep}{1pt}
   \caption{Statistically significant ANOVA effects ($p < 0.05$) across main, 2-way, and 3-way factor effects, over 25 tests for energy, latency, and \gls{ttft}, and 10 for accuracy.}
   \vspace{-2mm}
   \label{table:significant_tests}
   \resizebox{1\columnwidth}{!}{
   \begin{tabular}{lcccc}
      \toprule
      \textbf{Factor / Interaction}                                      %
                                                                 & \textbf{Energy}                      & \textbf{Latency} & \textbf{TTFT} & \textbf{Accuracy} \\
      \midrule
      Attention                                                  & 24                                   & 23               & 25            & 6                 \\
      Prefix Caching                                             & 18                                   & 21               & 22            & 6                 \\
      Chunked Prefill                                            & 2                                    & 1                & 1             & 1                 \\
      Attention $\times$ Prefix Caching                          & 4                                    & 10               & 11            & 3                 \\
      Attention $\times$ Chunked Prefill                         & 2                                    & 2                & 0             & 1                 \\
      Prefix Caching $\times$ Chunked Prefill                    & 2                                    & 1                & 1             & 1                 \\
      Attention $\times$ Prefix Caching $\times$ Chunked Prefill & 1                                    & 1                & 1             & 1                 \\
      \bottomrule
   \end{tabular}
   }
   \vspace{-3mm}
\end{table}

\subsubsection{\textbf{Hypothesis Testing}}
\citetable{significant_tests} shows counts of significant ANOVA tests ($p < 0.05$) for main effects, 2-way, and 3-way interactions across factors, using total energy, total latency, and \gls{ttft} metrics (out of $25$ total tests).
We observe a higher number of significant outcomes for conditions without an interaction effect of attention ($24$, $23$, and $25$) and prefix caching ($18$, $21$, and $22$), compared to the low numbers observed for chunked prefill ($2$, $1$, and $1$).
This observation suggests that attention and prefix caching have a greater influence on the dependent variables and that their interaction is noteworthy, particularly for total latency and \gls{ttft} ($10$ and $11$, respectively), but it is not dominant.
By contrast, effects involving chunked prefill are rare and are observed only on LB, the task with the longest prompts, which is consistent with the fact that shorter prompts are not expected to trigger prefill chunking under vLLM's default setup.

The interaction effects are inconsistent across models and tasks.
They are particularly frequent for \tech{Qwen-32B} and \tech{Llama-3B}, whereas no significant interaction is observed for \tech{Magistral-24B}.
LB has the highest concentration of interaction effects, including most of the significant three-way interactions, whereas NQ shows very few significant interactions overall.
These results show that interaction effects are model- and task-dependent rather than systematic.

\subsubsection{\textbf{Main effects of \gls{vllm} options}}
Table ~\ref{tab:posthoc-generation-energy-usage} reports ART ANOVA and Cliff's Delta tests on the direct impact of attention type, chunked prefill, and prefix caching on the energy consumption.
The same tests on performance and accuracy are available in the replication package~\cite{replicationpackage}. Non-significant effects are omitted from the post hoc analysis.

The results confirm that attention type has a significant impact on energy consumption across most models and metrics.
\tech{FlashInfer} generally has an advantage over \tech{FlashAttention-3}, with strong positive effect size across models, particularly for \tech{Qwen-4B} (AT and LB) and \tech{Llama-8B} (AT, LB, and WC).
\tech{FlashAttention-3}, in turn, tends to consume more than \tech{FlashAttention-2}.
Enabling prefix caching also has a significant impact, reducing energy consumption in almost all cases, even for tasks that are not normally affected by it. 
This reduction is particularly pronounced on EE, \tech{Magistral-24B}, and \tech{Qwen-32B}.
However, this trend is not universal: for \tech{Llama-3B} on AT, prefix caching leads to a negative effect size, indicating an increase in energy consumption in this specific case.
By contrast, enabling chunked prefill shows limited observable effects on energy consumption under our studied workloads.
As expected, the few significant cases are limited to LB, however, their effect sizes remain negligible.
The same overall trends are observed for total latency and \gls{ttft}, although the effects are less consistent than for energy consumption.
Attention type remains an important factor for both metrics, but the best-performing backend varies more across models and tasks than for energy.
Prefix caching has its clearest effect on \gls{ttft}, which it reduces in most cases; for total latency, the effect is weaker and more task-dependent.
Chunked prefill again shows limited observable effects under our studied workloads.

\begin{table}[t!]
\centering
\caption{Pairwise comparisons of inference options for Energy (kJ): $p_{Holm}$ and Cliff's $\delta$. 
AT: AssistantTraces, EE: EvoEval, LB: LongBench-v2, NQ: Natural Questions, WC: WildChat, v2: FlashAttention-2, v3: FlashAttention-3, inf: FlashInfer
}
\vspace{-2mm}
\label{tab:posthoc-generation-energy-usage}
\tiny
\setlength\tabcolsep{0.1pt}
\renewcommand\arraystretch{1}
\resizebox{1\columnwidth}{!}{
\begin{tabular}{lccccccc}
  \toprule
  \textbf{Options} & \textbf{Model} & \textbf{\shortstack{Contrast}} & \textbf{AT} & \textbf{EE} & \textbf{LB} & \textbf{NQ} & \textbf{WC} \\
  \midrule
  \multirow{15}{*}{\shortstack{Attention\\Type}} & \multirow{3}{*}{Q-32B} & f2 - fi & $\cellcolor[HTML]{FFCC99}<.001 (+.65)$ & $\cellcolor[HTML]{FFF2CC}<.001 (+.31)$ & $\cellcolor[HTML]{FFE699}<.001 (+.39)$ & $\cellcolor[HTML]{FFCC99}<.001 (+1.00)$ & $\cellcolor[HTML]{FFCC99}<.001 (+.71)$ \\
   &  & f2 - f3 & $\cellcolor[HTML]{FFCC99}<.001 (-.72)$ & $\cellcolor[HTML]{FFFFFF}.006 (+.09)$ & $\cellcolor[HTML]{FFF2CC}<.001 (+.22)$ & $\cellcolor[HTML]{FFCC99}<.001 (+1.00)$ & $\cellcolor[HTML]{FFFFFF}.087 (+.11)$ \\
   &  & f3 - fi & $\cellcolor[HTML]{FFCC99}<.001 (+.94)$ & $\cellcolor[HTML]{FFF2CC}<.001 (+.25)$ & $\cellcolor[HTML]{FFF2CC}.046 (+.26)$ & $\cellcolor[HTML]{FFCC99}<.001 (+.79)$ & $\cellcolor[HTML]{FFCC99}<.001 (+.63)$ \\
  \cmidrule(lr){2-8}
   & \multirow{3}{*}{Q-4B} & f2 - fi & $\cellcolor[HTML]{FFCC99}<.001 (+.74)$ & $\cellcolor[HTML]{FFCC99}<.001 (+.64)$ & $\cellcolor[HTML]{FFCC99}<.001 (+.93)$ & $\cellcolor[HTML]{FFCC99}<.001 (+1.00)$ & $\cellcolor[HTML]{FFCC99}<.001 (+.99)$ \\
   &  & f2 - f3 & $\cellcolor[HTML]{FFCC99}<.001 (-.90)$ & $\cellcolor[HTML]{FFCC99}<.001 (-.77)$ & $\cellcolor[HTML]{FFCC99}<.001 (-.62)$ & $\cellcolor[HTML]{FFCC99}<.001 (+.98)$ & $\cellcolor[HTML]{FFCC99}<.001 (-.50)$ \\
   &  & f3 - fi & $\cellcolor[HTML]{FFCC99}<.001 (+.98)$ & $\cellcolor[HTML]{FFCC99}<.001 (+.97)$ & $\cellcolor[HTML]{FFCC99}<.001 (+1.00)$ & $\cellcolor[HTML]{FFCC99}<.001 (+.88)$ & $\cellcolor[HTML]{FFCC99}<.001 (+1.00)$ \\
  \cmidrule(lr){2-8}
   & \multirow{3}{*}{M-24B} & f2 - fi & $\cellcolor[HTML]{FFF2CC}<.001 (+.22)$ & $\cellcolor[HTML]{FFF2CC}<.001 (+.21)$ & $\cellcolor[HTML]{FFFFFF}.441 (+.04)$ & $\cellcolor[HTML]{FFFFFF}.294 (+.06)$ & - \\
   &  & f2 - f3 & $\cellcolor[HTML]{FFFFFF}.515 (-.04)$ & $\cellcolor[HTML]{FFFFFF}.012 (-.11)$ & $\cellcolor[HTML]{FFCC99}<.001 (+1.00)$ & $\cellcolor[HTML]{FFF2CC}<.001 (+.30)$ & - \\
   &  & f3 - fi & $\cellcolor[HTML]{FFF2CC}<.001 (+.26)$ & $\cellcolor[HTML]{FFF2CC}<.001 (+.29)$ & $\cellcolor[HTML]{FFCC99}<.001 (-.99)$ & $\cellcolor[HTML]{FFF2CC}.019 (-.23)$ & - \\
  \cmidrule(lr){2-8}
   & \multirow{3}{*}{L-8B} & f2 - fi & $\cellcolor[HTML]{FFF2CC}<.001 (+.25)$ & $\cellcolor[HTML]{FFF2CC}<.001 (+.19)$ & $\cellcolor[HTML]{FFCC99}<.001 (+1.00)$ & $\cellcolor[HTML]{FFCC99}<.001 (+1.00)$ & $\cellcolor[HTML]{FFCC99}<.001 (+1.00)$ \\
   &  & f2 - f3 & $\cellcolor[HTML]{FFF2CC}<.001 (-.30)$ & $\cellcolor[HTML]{FFF2CC}<.001 (-.23)$ & $\cellcolor[HTML]{FFCC99}<.001 (-.58)$ & $\cellcolor[HTML]{FFF2CC}.006 (+.17)$ & $\cellcolor[HTML]{FFE699}<.001 (+.44)$ \\
   &  & f3 - fi & $\cellcolor[HTML]{FFCC99}<.001 (+.52)$ & $\cellcolor[HTML]{FFE699}<.001 (+.37)$ & $\cellcolor[HTML]{FFCC99}<.001 (+1.00)$ & $\cellcolor[HTML]{FFCC99}<.001 (+1.00)$ & $\cellcolor[HTML]{FFCC99}<.001 (+1.00)$ \\
  \cmidrule(lr){2-8}
   & \multirow{3}{*}{L-3B} & f2 - fi & $\cellcolor[HTML]{FFFFFF}.113 (-.08)$ & $\cellcolor[HTML]{FFF2CC}<.001 (+.30)$ & $\cellcolor[HTML]{FFCC99}<.001 (+.96)$ & $\cellcolor[HTML]{FFCC99}<.001 (+.75)$ & $\cellcolor[HTML]{FFCC99}<.001 (+1.00)$ \\
   &  & f2 - f3 & $\cellcolor[HTML]{FFF2CC}<.001 (-.17)$ & $\cellcolor[HTML]{FFCC99}<.001 (-.48)$ & $\cellcolor[HTML]{FFFFFF}.002 (-.09)$ & $\cellcolor[HTML]{FFCC99}<.001 (+.51)$ & $\cellcolor[HTML]{FFFFFF}.438 (+.04)$ \\
   &  & f3 - fi & $\cellcolor[HTML]{FFFFFF}.113 (+.13)$ & $\cellcolor[HTML]{FFCC99}<.001 (+.68)$ & $\cellcolor[HTML]{FFCC99}<.001 (+.89)$ & $\cellcolor[HTML]{FFE699}<.001 (+.34)$ & $\cellcolor[HTML]{FFCC99}<.001 (+1.00)$ \\
  \midrule
  \multirow{2}{*}{\shortstack{Chunked\\Prefill}} & \multirow{1}{*}{Q-32B} & off - on & - & - & $\cellcolor[HTML]{FFFFFF}.001 (-.04)$ & - & - \\
  \cmidrule(lr){2-8}
   & \multirow{1}{*}{L-3B} & off - on & - & - & $\cellcolor[HTML]{FFFFFF}.007 (+.07)$ & - & - \\
  \midrule
  \multirow{5}{*}{\shortstack{Prefix\\caching}} & \multirow{1}{*}{Q-32B} & off - on & $\cellcolor[HTML]{FFE699}<.001 (+.42)$ & $\cellcolor[HTML]{FFCC99}<.001 (+1.00)$ & $\cellcolor[HTML]{FFCC99}<.001 (+.93)$ & - & $\cellcolor[HTML]{FFE699}<.001 (+.47)$ \\
  \cmidrule(lr){2-8}
   & \multirow{1}{*}{Q-4B} & off - on & $\cellcolor[HTML]{FFF2CC}<.001 (+.20)$ & $\cellcolor[HTML]{FFE699}<.001 (+.44)$ & - & - & $\cellcolor[HTML]{FFF2CC}<.001 (+.25)$ \\
  \cmidrule(lr){2-8}
   & \multirow{1}{*}{M-24B} & off - on & $\cellcolor[HTML]{FFCC99}<.001 (+.60)$ & $\cellcolor[HTML]{FFCC99}<.001 (+.98)$ & - & - & $\cellcolor[HTML]{FFF2CC}<.001 (+.30)$ \\
  \cmidrule(lr){2-8}
   & \multirow{1}{*}{L-8B} & off - on & $\cellcolor[HTML]{FFCC99}<.001 (+.57)$ & $\cellcolor[HTML]{FFCC99}<.001 (+.69)$ & - & $\cellcolor[HTML]{FFFFFF}.003 (-.10)$ & $\cellcolor[HTML]{FFF2CC}<.001 (+.20)$ \\
  \cmidrule(lr){2-8}
   & \multirow{1}{*}{L-3B} & off - on & $\cellcolor[HTML]{FFF2CC}<.001 (-.26)$ & $\cellcolor[HTML]{FFCC99}<.001 (+.67)$ & $\cellcolor[HTML]{FFFFFF}.004 (-.09)$ & - & $\cellcolor[HTML]{FFF2CC}<.001 (+.19)$ \\
  \bottomrule
  \multicolumn{8}{l}{
        P-values lower than $\alpha = 0.05$ are shown in \textbf{bold}.
  } \\
  \multicolumn{8}{l}{
        Effect sizes: \marktext{efflarge}{ Large } - \marktext{effmedium}{ Medium } - \marktext{effsmall}{ Small } - \marktext{effnegligible}{ Negligible }
  }
\end{tabular}
}
\vspace{-2mm}
\end{table}

\subsubsection{\textbf{Interaction effects between \gls{vllm} options.}}

\begin{figure}[t]
   \centering
   \begin{subfigure}[t]{1\linewidth}
      \centering
      \begin{subfigure}[t]{0.49\linewidth}
         \centering
         \includegraphics[width=\linewidth]{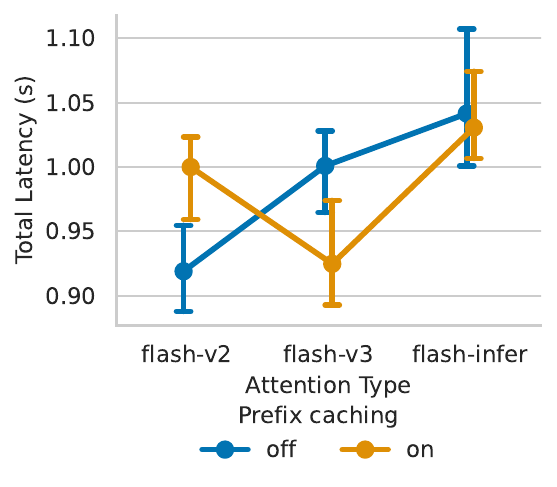}
         \vspace{-4mm}
         \caption{Llama-3B-Instruct with AssistantTraces}
         \label{fig:2_way_interaction_llama}
      \end{subfigure}
      \hfill
      \begin{subfigure}[t]{0.49\linewidth}
         \centering
         \includegraphics[width=\linewidth]{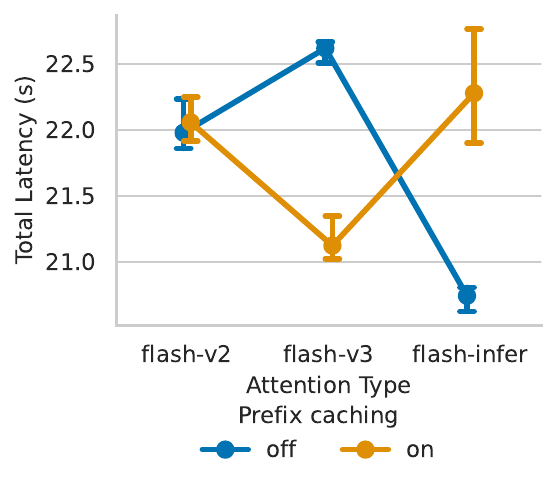}
         \vspace{-4mm}
         \caption{Qwen-4B with LongBench}
         \label{fig:2_way_interaction_qwen3}
      \end{subfigure}
      \label{fig:2_way_interactions_latency}
   \end{subfigure}
   \hfill
   
   \caption{Representative 
   2-way interaction effects on end-to-end request latency.}
   \label{fig:interactions_latency}
   \vspace{-3mm}
\end{figure}

\citefig{interactions_latency} shows representative 2-way interaction patterns observed for end-to-end request latency.
In \citefig{2_way_interaction_llama} and \citefig{2_way_interaction_qwen3}, enabling prefix caching decreases latency when \tech{FlashAttention-3} is used, for both \tech{Llama-3B} and \tech{Qwen-4B}. 
This pattern is not limited to these two examples: with \tech{FlashAttention-3}, prefix caching improves latency in 37/50 cases, \gls{ttft} in 38/50 cases, and generation energy usage in 40/50 cases. 
\tech{FlashAttention-3} is the backend for which prefix caching appears slightly more consistently beneficial. 
Still, similar gains are also observed with \tech{FlashAttention-2} and \tech{FlashInfer}: for latency, prefix caching improves 29/50 cases with \tech{FlashAttention-2} and 31/50 with \tech{FlashInfer}; for \gls{ttft}, 34/50 with \tech{FlashAttention-2} and 35/50 with \tech{FlashInfer}; and for generation energy usage, 36/50 with \tech{FlashAttention-2} and 37/50 with \tech{FlashInfer}. %
The effect is not exclusive to \tech{FlashAttention-3}, but only slightly more consistent with it.

\subsubsection{\textbf{Anomalous effects of our factors on accuracy}}
We identified statistically significant effects of inference configuration parameters, especially attention type and prefix caching, on benchmark accuracy scores across several model--benchmark pairs (\citetable{significant_tests}), with chunked prefill showing only one significant test of negligible effect.
These findings are unexpected, as the parameters in question are intended solely for computational optimization and have no bearing on model outputs.

\textbf{Effect of attention type.}
The clearest case is \tech{Qwen-4B} on LB, where \tech{FlashInfer} reached accuracy in the $0.37$--$0.43$ range against $0.20$--$0.35$ for both \tech{FlashAttention} variants, differing from each with a large effect, while the two \tech{FlashAttention} variants differed only slightly.
The same pattern appears more weakly on EE, where \tech{FlashInfer} significantly outperforms \tech{FlashAttention-2} for all models (small-to-medium effects) and \tech{FlashAttention-3} for all but \tech{Magistral-24B}.
The exception is \tech{Llama-3B} on LB, where \tech{FlashAttention-2} performs worse than both other backends, which are themselves indistinguishable.

\textbf{Effect of prefix caching.}
Enabling prefix caching significantly altered measured accuracy for six model--benchmark combinations.
It reduced accuracy on EE for \tech{Magistral-24B} (large effect) and, more weakly, \tech{Llama-8B}, but increased it on EE for \tech{Qwen-32B}, \tech{Qwen-4B}, and \tech{Llama-3B} (largest positive shift for the latter).
\tech{Llama-3B} on LB also differed significantly but with negligible effect size.
The exact statistics for all contrasts are reported in the replication package~\cite{replicationpackage}.

\begin{result}
    \textbf{RQ1:} Attention type and prefix caching show the most frequent effects on energy and performance. \tech{FlashInfer} often reduces energy, while the fastest backend depends on the settings. Prefix caching mainly affects \gls{ttft}, whereas chunked prefill has limited effects even on the long-prompt task. We also observe significant accuracy differences for some model--benchmark pairs. \noindent \textbf{Implication:} attention type and prefix caching should be evaluated jointly across energy, performance, and accuracy.
\end{result}

\subsection{Interaction Effects between Configurations \& Task Types (RQ2)}\label{subsec:rq2}
\subsubsection{\textbf{Data exploration}}
\begin{figure}[t]
   \centering
   \includegraphics[width=\linewidth]{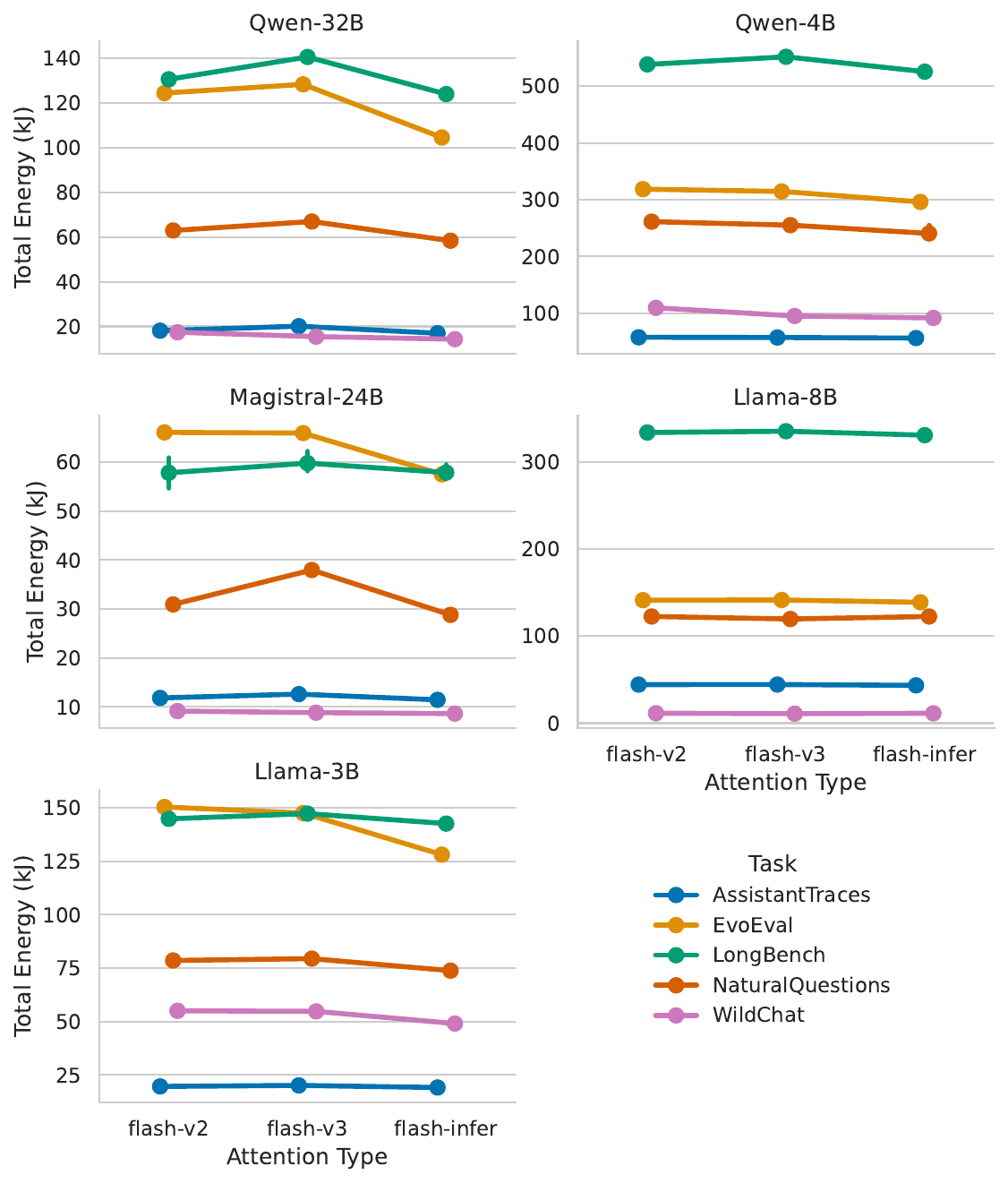}
   \caption{Task and attention type interaction on total energy consumption (kJ)}
\label{fig:interaction_workload_attn_generation_energy_usage}
\vspace{-6mm}
\end{figure}

{\textbf{\textit{Interaction between task and attention}}}.
As shown in \Cref{fig:interaction_workload_attn_generation_energy_usage}, the task is the dominant factor influencing energy consumption, introducing a stronger effect than the attention parameter in every case.
\tech{FlashInfer} tends to reduce energy usage compared to \tech{FlashAttention-2} and \tech{FlashAttention-3} across most models and tasks. 
However, in certain configurations, such as specific models or tasks like EE, the attention type has minimal impact. %
Model size also plays a substantial role, with larger models consistently consuming more energy.
Latency and \gls{ttft} have related, but not identical, patterns (the corresponding figures are provided in the replication package).
In several cases, \tech{FlashAttention-3} yields lower latency and slightly lower \gls{ttft} than \tech{FlashAttention-2}, showing that although \tech{FlashInfer} often achieves the lowest energy consumption, it is not systematically the fastest option.
This suggests a trade-off between time and energy efficiency, as the attention type that minimizes energy consumption is not always the one that minimizes response time.

\begin{figure}[t]
   \centering
   \includegraphics[width=1\linewidth]{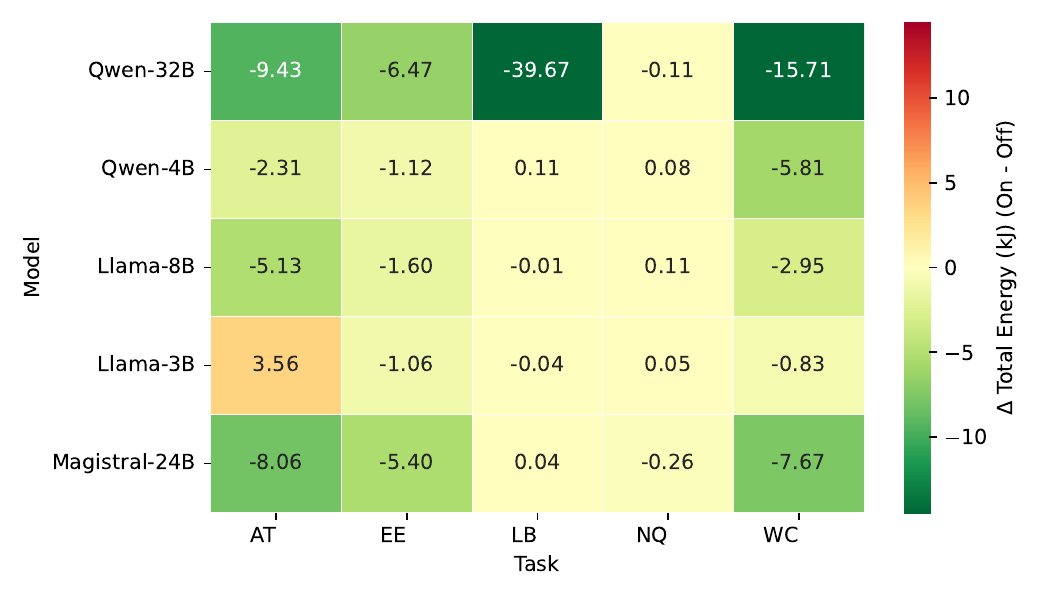}
    \caption{Difference in total energy consumption (prefix caching enabled minus disabled) across tasks and models, in\,kJ.} 
    \vspace{-6mm}
   \label{fig:significant_workload_x_prefix_caching_heatmap}
\end{figure}

\textbf{Interaction between task and prefix caching.}~\Cref{fig:significant_workload_x_prefix_caching_heatmap} summarizes this interaction for total energy consumption by showing, for each model and task pair, the change in median total energy when prefix caching is enabled rather than disabled (\(\Delta = \textit{On} - \textit{Off}\)).
Negative values, therefore, indicate lower energy consumption with prefix caching, whereas positive values indicate an increase.

\autoref{fig:significant_workload_x_prefix_caching_heatmap} shows that, despite a statistically significant interaction, enabling prefix caching generally leads to modest decreases in energy consumption, with effects that remain strongly task-dependent. The largest decreases

Latency and \gls{ttft} show a similar overall pattern: their interactions with the task are also statistically significant, but the corresponding changes induced by prefix caching remain limited across most model-task combinations.
In most cases, enabling prefix caching slightly reduces latency and \gls{ttft}, although some combinations show almost no effect and a few exhibit small increases.
Runtime metrics in the replication package~\cite{replicationpackage}, including prefix-cache queries and hits, confirm that benefits depend on workload characteristics and cache utilization.

Overall, task remains the dominant factor, while prefix caching introduces a secondary, statistically detectable effect that should be interpreted as a small task-dependent optimization rather than a uniform benefit.

\subsubsection{\textbf{Hypothesis Testing}}
The ART ANOVA results revealed a significant interaction between attention type and task, and between prefix caching and task, in determining total energy, latency, and \gls{ttft} across all tested models (15 out of 15 tests were significant for each factor).
By contrast, chunked prefill shows no significant task-level interaction (0/15), even with LB included as the long-context workload

\begin{result}
\textbf{RQ2:}
Attention type and prefix caching show significant interactions with the task, while chunked prefill does not. Attention type leads to substantial task-dependent variations, often revealing trade-offs between energy and latency. In contrast, prefix caching yields smaller and less consistent improvements.
\noindent \textbf{Implication:} configuration effects are heavily task-dependent and must be evaluated on representative datasets rather than in isolation.
\end{result}

\subsection{Pareto-optimal \gls{vllm} Configurations (RQ3)}\label{subsec:rq3}

\begin{figure}[t!]
   \centering
   \includegraphics[width=1\linewidth]{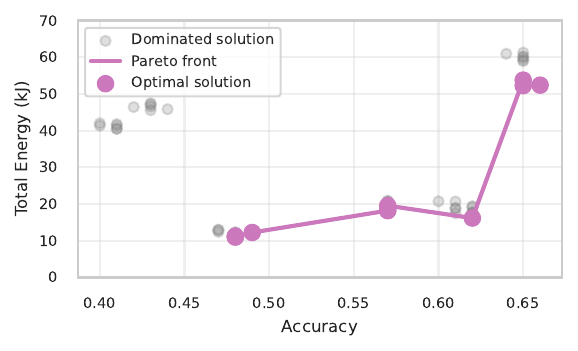}
   \vspace{-3mm}
    \caption{Two-dimensional projection of the EvoEval Pareto front over energy (kJ) and accuracy across configurations and selected \glspl{llm}. The front is computed over energy, latency, and accuracy; latency is reported in Table \ref{tab:pareto-optimal-configs}.}
   \vspace{-4mm}
   \label{fig:pareto-ee}
\end{figure}

The Pareto analysis confirms that no single \gls{vllm} configuration is optimal across all tasks. 
For each task, the Pareto front is computed over latency, energy consumption, and accuracy when available. 
For readability, Figure~\ref{fig:pareto-ee} shows a two-dimensional projection of this front, focusing on energy and accuracy, while the full Pareto-optimal configurations, including latency, are reported in Table~\ref{tab:pareto-optimal-configs}.
On EE (Figure~\ref{fig:pareto-ee}), the Pareto front is mainly driven by the choice of model rather than by \gls{vllm} options.
\tech{Llama-3B} occupies the low-cost end of the front (with energy between 11.0 and 12.2\,kJ, latency between 1.61 and 1.71\,s, and accuracy between 0.48 and 0.49). 
In contrast, \tech{Qwen-4B} provides a more balanced trade-off, increasing accuracy to 0.62 (+27\% compared to the best \tech{Llama-3B} point), at the cost of a 33--47\% increase in energy and a 63--73\% increase in latency (16.2\,kJ and 2.78\,s).
\tech{Llama-8B} lies between these regimes, reaching 0.57 accuracy with 18.2--19.5\,kJ and 2.48--2.63\,s, whereas \tech{Qwen-32B} achieves the highest accuracy (0.65--0.66) but at the cost of a 223--232\% increase in energy and a 62--64\% increase in latency compared to \tech{Qwen-4B}.
Within a given model, \gls{vllm} parameters mainly move configurations locally along the front.
For instance, for \tech{Qwen-32B}, switching from \tech{FlashAttention-3} to \tech{FlashInfer} reduces energy consumption by about 3\% with negligible impact on accuracy. In contrast, for \tech{Llama-8B}, \tech{FlashInfer} reduces energy by about 7\% at the cost of a small latency increase (5--6\%).
Similar patterns are observed across other tasks.
As detailed in Table~\ref{tab:pareto-optimal-configs}, on LB all Pareto-optimal points use \tech{Qwen-4B} with \tech{FlashInfer}, trading latency and energy for accuracy.
On NQ, the Pareto set reflects a small latency--energy trade-off, with the lowest latency obtained from \tech{FlashAttention-3} (no prefix caching or chunked prefill) and the lowest energy from \tech{FlashInfer} with chunked prefill.
On WC, \tech{FlashInfer} consistently reduces energy compared with \tech{FlashAttention-3} at the cost of a small latency increase.
Finally, AT has a single optimal configuration: \tech{Llama-3B} with \tech{FlashAttention-2} and chunked prefill.

Overall, these results show that model choice determines the main performance regime, while 
\gls{vllm} configuration options (attention kernel, prefix caching, chunked prefill) provide fine-grained adjustments within that regime.

\begin{result}
\textbf{RQ3:}
No single configuration is optimal across tasks.
Model choice dominates the Pareto front, defining the main trade-offs between energy, latency, and accuracy.
\gls{vllm} configuration options only induce local adjustments, typically yielding modest energy savings or latency trade-offs without changing the overall regime. \noindent \textbf{Implication:} selecting the right model is the primary decision, while configuration tuning provides secondary, fine-grained optimization whose benefits are task-dependent.
\end{result}

\begin{table}[ht]
   \centering
   \tiny
   \setlength\tabcolsep{2.5pt}
   \renewcommand\arraystretch{1.03}
   \caption{Pareto-optimal configurations by task. Only frontier models are shown. Best values are highlighted in green, worst in red. 
   AT: AssistantTraces, EE: EvoEval, LB: LongBench-v2, NQ: Natural Questions, WC: WildChat,
   v2: FlashAttention-2, v3: FlashAttention-3, infer: FlashInfer
   }
   \vspace{-2mm}
   \label{tab:pareto-optimal-configs}
   \resizebox{0.98\columnwidth}{!}{
   \begin{tabular}{l l c c c r r r}
      \toprule
      \textbf{\shortstack{Task\\Type}}             & \textbf{Model}             & \textbf{\shortstack{Attention\\Type}} & \textbf{\shortstack{Prefix\\Caching}} & \textbf{\shortstack{Chunked\\Prefill}} & \textbf{Energy (kJ)}     & \textbf{Latency (s)}    & \textbf{Accuracy}         \\
      \midrule
      \multirow{1}{*}{\textbf{AT}}  & \multirow{1}{*}{Llama-3B}  & f2           & $\times$                & \checkmark               & 46.0                     & 0.910                      & -                         \\
      \addlinespace
      \midrule
      \multirow{11}{*}{\textbf{EE}} & \multirow{3}{*}{Qwen-32B} & f3           & \checkmark              & $\times$                 & \cellcolor{red!20}53.8   & \cellcolor{green!20}4.490  & \cellcolor{red!20}0.650   \\
                                    &                            & fi        & \checkmark              & $\times$                 & \cellcolor{green!20}52.3 & 4.520                      & \cellcolor{red!20}0.650   \\
                                    &                            & fi        & \checkmark              & \checkmark               & 52.4                     & \cellcolor{red!20}4.560    & \cellcolor{green!20}0.660 \\
      \addlinespace
                                    & \multirow{1}{*}{Qwen-4B}  & fi        & \checkmark              & \checkmark               & 16.2                     & 2.780                      & 0.620                     \\
      \addlinespace
                                    & \multirow{3}{*}{Llama-8B}  & f3           & \checkmark              & $\times$                 & \cellcolor{red!20}19.5   & \cellcolor{green!20}2.480  & \cellcolor{green!20}0.570 \\
                                    &                            & fi        & \checkmark              & $\times$                 & 18.2                     & 2.610                      & \cellcolor{green!20}0.570 \\
                                    &                            & fi        & \checkmark              & \checkmark               & \cellcolor{green!20}18.2 & \cellcolor{red!20}2.630    & \cellcolor{green!20}0.570 \\
      \addlinespace
                                    & \multirow{4}{*}{Llama-3B}  & f2           & \checkmark              & \checkmark               & 11.2                     & \cellcolor{green!20}1.610  & \cellcolor{red!20}0.480   \\
                                    &                            & f3           & \checkmark              & \checkmark               & \cellcolor{red!20}12.2   & 1.660                      & \cellcolor{green!20}0.490 \\
                                    &                            & fi        & \checkmark              & \checkmark               & 11.1                     & 1.700                      & \cellcolor{red!20}0.480   \\
                                    &                            & fi        & \checkmark              & $\times$                 & \cellcolor{green!20}11.0 & \cellcolor{red!20}1.710    & \cellcolor{red!20}0.480   \\
      \addlinespace
      \midrule
      \multirow{4}{*}{\textbf{LB}}  & \multirow{3}{*}{Qwen-4B}  & fi        & $\times$                & \checkmark               & \cellcolor{green!20}56.8 & \cellcolor{green!20}20.730 & \cellcolor{red!20}0.370   \\
                                    &                            & fi        & \checkmark              & $\times$                 & 60.3                     & 21.900                     & 0.410                     \\
                                    &                            & fi        & \checkmark              & \checkmark               & \cellcolor{red!20}61.3   & \cellcolor{red!20}22.630   & \cellcolor{green!20}0.420 \\
      \addlinespace
      \midrule
      \multirow{3}{*}{\textbf{NQ}}  & \multirow{3}{*}{Llama-3B}  & f3           & $\times$                & $\times$                 & \cellcolor{red!20}8.8    & \cellcolor{green!20}3.240  & -                         \\
                                    &                            & f3           & \checkmark              & \checkmark               & 8.8                      & 3.310                      & -                         \\
                                    &                            & fi        & $\times$                & \checkmark               & \cellcolor{green!20}8.6  & \cellcolor{red!20}3.580    & -                         \\
      \addlinespace
      \midrule
      \multirow{4}{*}{\textbf{WC}}  & \multirow{4}{*}{Llama-3B}  & f3           & $\times$                & $\times$                 & \cellcolor{red!20}66.0   & \cellcolor{green!20}6.680  & -                         \\
                                    &                            & f3           & \checkmark              & $\times$                 & 65.1                     & 6.810                      & -                         \\
                                    &                            & fi        & $\times$                & $\times$                 & 57.9                     & 6.900                      & -                         \\
                                    &                            & fi        & \checkmark              & $\times$                 & \cellcolor{green!20}56.7 & \cellcolor{red!20}6.920    & -                         \\
      \addlinespace
      \bottomrule
   \end{tabular}
   }
   \vspace{-4mm}
\end{table}

\section{Discussion}
\label{sec:discussion-and-limitations}

This section contextualizes the results in \Cref{sec:results}.
Across the three RQs, no \gls{vllm} configuration dominates across all models and tasks: optimal settings are model- and task-specific. Attention kernel and prefix caching are the main levers, while chunked prefill has negligible effects in our settings.
Overall, \gls{llm} and task drive most variance, with configurations acting as secondary but non-negligible modifiers.
Unexpectedly, inference-time optimizations intended to be output-neutral can measurably affect benchmark accuracy, with implications for \gls{llm} evaluation reproducibility.

\paragraph{\textbf{Configuration Effects on Benchmark Accuracy}}
Prefix caching and attention backend selection are inference-time optimizations that operate exclusively at the computational level: they should not alter either model weights or the mathematical operations they approximate.
In principle, then, the change in benchmark accuracy under these settings should be zero. Yet we observe non-zero shifts. Two explanations are possible: the differences are statistical noise from a finite evaluation set, or they reflect a real effect---for example, attention kernels and cache code paths reorder floating-point reductions, and floating-point addition is non-associative, so numerically distinct (though mathematically equivalent) results can propagate into different sampled tokens. Disentangling these would require repeated runs with fixed seeds and per-token logit comparison, which we leave to future work.

These results imply that published benchmark scores may not be reproducible across inference stacks when the full configuration is not reported, although further research is needed to correctly assess the effect.
If this effect is not a statistical anomaly, practitioners should not assume that a model's published benchmark performance will transfer to a different deployment configuration, and empirical studies evaluating \gls{llm} accuracy should report their full inference setup as part of their experimental protocol.

\paragraph{\textbf{Interpreting Cross-cutting Patterns}}

\textit{Why chunked prefill had negligible effects.}
The almost nonexistent effect of the prefill chunk can be explained by the interaction between our task characteristics and the default value of vLLM's option \texttt{max\_num\_batched\_tokens} of 8,192.
A prompt is only chunked when it exceeds this token budget. 
Four of our five tasks have average prompt lengths well below this threshold, meaning that chunked prefill was never actually triggered for the vast majority of requests in these tasks. Furthermore, vLLM's documentation states that 8,192 is the recommended value for optimizing throughput, which directly affects total energy and end-to-end latency.
Conversely, setting \texttt{max\_num\_batched\_tokens} too low degrades performance, since small chunks reduce the compute intensity of each prefill step and leave the GPU underutilized~\cite{vllmOptimizationDocs}.
Our default-budget configuration therefore reflects the regime vLLM recommends, and the negligible effect we observe is specific to that recommended setting rather than to chunked prefill in general.

\textit{Why energy and latency are not proportional.}
A practically important and recurring observation is that \tech{FlashInfer} tends to achieve the lowest energy consumption while \tech{FlashAttention-3} tends to achieve the lowest latency.
This decoupling highlights that optimizing for energy and optimizing for latency can require different configuration choices, and that reporting only one metric is insufficient for a complete characterization of efficiency.
Isolating the contribution of individual task-level factors through fine-grained profiling would be a valuable direction for follow-up work, as would measuring per-configuration memory bandwidth utilization.

\textit{KV cache block allocation and the LB prefix caching anomaly.}
For \tech{Qwen-32B} on LB, enabling prefix caching reduces energy with a large effect ($\delta = 0.93$), yet LB contains no repeated prompt prefixes, ruling out cache hits as the cause.
A plausible explanation is that enabling prefix caching alters vLLM's KV-cache block allocation and memory-management path even when no prefix is reused, changing GPU memory access patterns independently of cache hits.
Confirming this would require instrumenting block-level allocation and reproducing the effect on additional long-context datasets, which we leave to future work.

\paragraph{\textbf{Implications for Stakeholders.}}
\textit{Prioritize attention kernel and prefix caching; chunked prefill can be left at its default.}
Among the three options studied, attention kernel and prefix caching are the dominant configuration levers.
\tech{FlashInfer} should be the default choice when energy efficiency is the priority, while \tech{FlashAttention-3} is preferable when minimizing latency.
Prefix caching reduces energy and latency in most cases, but practitioners should validate this on their specific task.
Chunked prefill, by contrast, had no meaningful effect across our experiment and can safely be left at its default value.

\textit{Use the Pareto frontier as a decision tool.}
The Pareto-optimal configurations in \Cref{tab:pareto-optimal-configs} provide practitioners with a principled basis for selecting configurations under multi-objective constraints.
Rather than picking the configuration that ranks best on a single metric, the frontier identifies the set of configurations where no alternative is strictly better on \emph{all} objectives simultaneously.
In practice, this means: if energy efficiency is the primary concern, select the Pareto-optimal point with the lowest energy value (typically \tech{FlashInfer} with prefix caching enabled); if latency is critical, select the point with the lowest latency (often \tech{FlashAttention-3} with prefix caching enabled).

\textit{Evaluate options jointly.}
The attention $\times$ prefix caching interaction is significant in 10/25 latency and 11/25 \gls{ttft} tests---i.e., the best attention kernel may depend on whether prefix caching is enabled.
Options evaluated independently can lead to suboptimal choices.
We recommend that practitioners / future benchmarking efforts adopt a factorial design to capture such interactions, and that vendors consider task-aware defaults over a single static configuration.

\textit{For researchers: report the full inference configuration.}
Our results show that attention backend and prefix caching settings can shift accuracy scores by margins comparable to differences between models of different sizes.
Benchmark results reported without specifying the full configuration of the inference engine are therefore not fully reproducible: a replication using a different inference stack may produce substantially different accuracy scores for the same model.
In line with \etal{Baltes} guidelines~\cite{baltesGuidelinesEmpiricalStudies2025}, we recommend that empirical studies evaluating \gls{llm} accuracy include the complete inference configuration in their protocol.
\section{Threats to Validity}
\label{sec:threats}
\textbf{Construct Validity.}
We measure GPU and CPU energy using \texttt{nvidia-smi} and \texttt{perf}, respectively. On A100 GPUs, \texttt{nvidia-smi} samples power for only 25\% of runtime, interpolating the rest, which can cause energy errors of up to 65\% in very spiky workloads~\cite{yangParttimePowerMeasurements2024}. 
Repeating runs reduces noise but not systematic bias, energy values should therefore be interpreted as estimates.
Accuracy was evaluated only for EE and LB, as AT, WC, and NQ lack suitable ground truth. These benchmarks may not represent accuracy across all task types.
With sampling-based decoding, accuracy differences may also partly reflect stochastic generation effects 
The 30 repetitions help capture this variability while preserving a realistic setup.

\textbf{Internal Validity.}
To keep the study tractable, we evaluated three \gls{vllm} configurations while holding all other settings at their defaults. These defaults may interact with the studied factors in unseen ways, but exhaustive exploration is infeasible~\cite{zine2026pimp}. We therefore focus on practically relevant options likely to act independently. In addition, under \gls{vllm}'s default setup, enabling chunked prefill does not necessarily activate the mechanism for short-prompt tasks. Therefore, for such tasks, our results capture the effect of enabling the option rather than the behavior of active prefill chunking.

\textbf{External Validity.}
Our experiments used NVIDIA A100-SXM4-40GB GPUs. Since attention kernel performance depends on hardware, \eg \tech{FlashAttention-3} is tuned for H100 GPUs, results on A100s may not generalize to other platforms.
Our results reflect a specific setup using \gls{vllm} 0.10.2, five models from three families, and five datasets. Because inference optimizations evolve quickly, and our models and workloads cover only part of the design space, generalization to other versions, architectures, scales, and tasks should be made with caution.
We use \gls{vllm}'s offline batch inference interface rather than its HTTP server mode, where asynchronous requests, scheduling, and cache behavior may differ. Finally, our findings are specific to \gls{vllm}; other inference servers %
may implement similar options differently.

\textbf{Conclusion Validity.}
Although $30$ repetitions per configuration is substantial compared to prior work, it may still be limited given the variability of \gls{llm} inference. Energy measurements were highly stable (\text{CV} = 0.043). To limit false positives, we applied the Holm–Bonferroni correction~\cite{holm1979simple} and used ART ANOVA for our non-normal data~\cite{wobbrock2011aligned}.
\section{Conclusion}
\label{sec:conclusion}

Rather than providing universal deployment guidelines, we presented a controlled exploratory study of how three \gls{vllm} options---attention kernel, prefix caching, and chunked prefill---affect energy consumption, performance, and accuracy across five \glspl{llm} and five tasks. We identify trade-offs that deserve further investigation and call for treating the inference stack as a first-class experimental variable.

The broader lesson is that \gls{vllm} configuration is not a neutral implementation detail. Defaults or unreported choices can affect energy, latency, and accuracy conclusions. Navigating this space requires a task-aware, multi-objective perspective, as evaluating options in isolation can lead to suboptimal decisions.

Future work should extend this factorial design to other \gls{vllm} parameters, hardware settings, multi-GPU deployments, and MoE architectures. It should also further investigate chunked prefill on workloads where prefill chunking is systematically triggered.

\textbf{Data Availability Statement:}
All artifacts (code, datasets, and results) are made available in an online replication package~\cite{replicationpackage}.

\newpage
\bibliographystyle{IEEEtran}
\balance
\bibliography{bibfile.bib}

@inproceedings{yangParttimePowerMeasurements2024,
  title = {Part-Time {{Power Measurements}}: Nvidia-Smi's {{Lack}} of {{Attention}}},
  shorttitle = {Part-Time {{Power Measurements}}},
  booktitle = {{{SC24}}: {{International Conference}} for {{High Performance Computing}}, {{Networking}}, {{Storage}} and {{Analysis}}},
  author = {Yang, Zeyu and Adamek, Karel and Armour, Wesley},
  year = 2024,
  month = nov,
  eprint = {2312.02741},
  primaryclass = {cs},
  pages = {1--17},
  doi = {10.1109/SC41406.2024.00028},
  urldate = {2025-03-20},
  archiveprefix = {arXiv}
}

@article{vries-gao_recalibrating_2026,
	title = {Recalibrating global artificial intelligence e-waste estimates},
	volume = {229},
	issn = {0921-3449},
	url = {https://www.sciencedirect.com/science/article/pii/S0921344926000960},
	doi = {https://doi.org/10.1016/j.resconrec.2026.108872},
	journal = {Resources, Conservation and Recycling},
	author = {Vries-Gao, Alex de},
	year = {2026},
	pages = {108872},
}

@article{jegham2025hungry,
  title={How hungry is ai? benchmarking energy, water, and carbon footprint of llm inference},
  author={Jegham, Nidhal and Abdelatti, Marwan and Koh, Chan Young and Elmoubarki, Lassad and Hendawi, Abdeltawab},
  journal={arXiv preprint arXiv:2505.09598},
  year={2025}
}

@inproceedings{ding2024sustainable,
  title={Sustainable LLM serving: Environmental implications, challenges, and opportunities},
  author={Ding, Yi and Shi, Tianyao},
  booktitle={2024 IEEE 15th International Green and Sustainable Computing Conference (IGSC)},
  pages={37--38},
  year={2024},
  organization={IEEE}
}

@article{basili1988tame,
	title        = {The TAME project: Towards improvement-oriented software environments},
	author       = {Basili, Victor R and Rombach, H Dieter},
	year         = 1988,
	journal      = {IEEE Transactions on Software Engineering},
	publisher    = {IEEE},
	volume       = 14,
	number       = 6,
	pages        = {758--773}
}

@misc{replicationpackage,
	title = {Replication package of this study},
	url = {https://doi.org/10.5281/zenodo.19253906},
	publisher = {Zenodo},
	author = {Anonymous},
	month = aug,
	year = {2026},
	doi = {https://doi.org/10.5281/zenodo.19253906},
}

@article{FGCS_2024,
  author = {Achim Guldner and {\em et al.}},
  doi = { https://doi.org/10.1016/j.future.2024.01.033 },
  issn = { 0167-739X },
  year = { 2024 },
  pages = { 402-418 },
  volume = { 155 },
  journal = { Future Generation Computer Systems },
  title = { Development and evaluation of a reference measurement model for assessing the resource and energy efficiency of software products and components-Green Software Measurement Model (GSMM) },
}

@book{wohlin2012experimentation,
	title        = {Experimentation in Software Engineering},
	author       = {Wohlin, C. and Runeson, P. and H{\"o}st, M. and Ohlsson, M.C. and Regnell, B. and Wessl{\'e}n, A.},
	year         = 2012,
	publisher    = {Springer},
	series       = {Computer Science},
	isbn         = 9783642290442
}

@book{shull2007guide,
	title        = {Guide to advanced empirical software engineering},
	author       = {Shull, Forrest and Singer, Janice and Sj{\o}berg, Dag IK},
	year         = 2007,
	publisher    = {Springer}
}

@article{murtaza2025impact,
  title={The impact of LLM chatbots on learning outcomes in advanced driver assistance systems education},
  author={Murtaza, Mohsin and Cheng, Chi-Tsun and Albahlal, Bader M and Muslam, Muhana Magboul Ali and Raza, Mansoor Syed},
  journal={Scientific Reports},
  volume={15},
  number={1},
  pages={7260},
  year={2025},
  publisher={Nature Publishing Group UK London}
}

@article{bharathi2024analysis,
  title={An analysis of large language models: their impact and potential applications},
  author={Bharathi Mohan, G and Prasanna Kumar, R and Vishal Krishh, P and Keerthinathan, A and Lavanya, G and Meghana, Meka Kavya Uma and Sulthana, Sheba and Doss, Srinath},
  journal={Knowledge and Information Systems},
  volume={66},
  number={9},
  pages={5047--5070},
  year={2024},
  publisher={Springer}
}

@article{eloundou2024gpts,
  title={GPTs are GPTs: Labor market impact potential of LLMs},
  author={Eloundou, Tyna and Manning, Sam and Mishkin, Pamela and Rock, Daniel},
  journal={Science},
  volume={384},
  number={6702},
  pages={1306--1308},
  year={2024},
  publisher={American Association for the Advancement of Science}
}

@article{park2025survey,
  title={A Survey on Inference Engines for Large Language Models: Perspectives on Optimization and Efficiency},
  author={Park, Sihyeong and Jeon, Sungryeol and Lee, Chaelyn and Jeon, Seokhun and Kim, Byung-Soo and Lee, Jemin},
  journal={preprint arXiv:2505.01658},
  year={2025}
}

@inproceedings{kwon2023efficient,
  title={Efficient memory management for large language model serving with pagedattention},
  author={Kwon, Woosuk and Li, Zhuohan and Zhuang, Siyuan and Sheng, Ying and Zheng, Lianmin and Yu, Cody Hao and Gonzalez, Joseph and Zhang, Hao and Stoica, Ion},
  booktitle={Proceedings of the 29th symposium on operating systems principles},
  pages={611--626},
  year={2023}
}

@article{fernandez2025energy,
  title={Energy considerations of large language model inference and efficiency optimizations},
  author={Fernandez, Jared and Na, Clara and Tiwari, Vashisth and Bisk, Yonatan and Luccioni, Sasha and Strubell, Emma},
  journal={preprint arXiv:2504.17674},
  year={2025}
}

@inproceedings{patel2024characterizing,
  title={Characterizing power management opportunities for llms in the cloud},
  author={Patel, Pratyush and Choukse, Esha and Zhang, Chaojie and Goiri, {\'I}{\~n}igo and Warrier, Brijesh and Mahalingam, Nithish and Bianchini, Ricardo},
  booktitle={Proceedings of the 29th ACM International Conference on Architectural Support for Programming Languages and Operating Systems, Volume 3},
  pages={207--222},
  year={2024}
}

@article{stojkovic2024towards,
  title={Towards greener llms: Bringing energy-efficiency to the forefront of llm inference},
  author={Stojkovic, Jovan and Choukse, Esha and Zhang, Chaojie and Goiri, Inigo and Torrellas, Josep},
  journal={arXiv preprint arXiv:2403.20306},
  year={2024}
}

@article{wilkins2024offline,
  title={Offline energy-optimal llm serving: Workload-based energy models for llm inference on heterogeneous systems},
  author={Wilkins, Grant and Keshav, Srinivasan and Mortier, Richard},
  journal={ACM SIGENERGY Energy Informatics Review},
  volume={4},
  number={5},
  pages={113--119},
  year={2024},
  publisher={ACM New York, NY, USA}
}

@inproceedings{luccioni2024power,
  title={Power hungry processing: Watts driving the cost of AI deployment?},
  author={Luccioni, Sasha and Jernite, Yacine and Strubell, Emma},
  booktitle={Proceedings of the 2024 ACM conference on fairness, accountability, and transparency},
  pages={85--99},
  year={2024}
}

@article{maliakel2025investigating,
  title={Investigating Energy Efficiency and Performance Trade-offs in LLM Inference Across Tasks and DVFS Settings},
  author={Maliakel, Paul Joe and Ilager, Shashikant and Brandic, Ivona},
  journal={arXiv preprint arXiv:2501.08219},
  year={2025}
}

@inproceedings{stojkovic2025dynamollm,
  title={Dynamollm: Designing llm inference clusters for performance and energy efficiency},
  author={Stojkovic, Jovan and Zhang, Chaojie and Goiri, {\'I}{\~n}igo and Torrellas, Josep and Choukse, Esha},
  booktitle={2025 IEEE International Symposium on High Performance Computer Architecture (HPCA)},
  pages={1348--1362},
  year={2025},
  organization={IEEE}
}

@article{husom2025sustainable,
  title={Sustainable llm inference for edge ai: Evaluating quantized llms for energy efficiency, output accuracy, and inference latency},
  author={Husom, Erik Johannes and Goknil, Arda and Astekin, Merve and Shar, Lwin Khin and K{\aa}sen, Andre and Sen, Sagar and Mithassel, Benedikt Andreas and Soylu, Ahmet},
  journal={arXiv preprint arXiv:2504.03360},
  year={2025}
}

@inproceedings{samsi2023words,
  title={From words to watts: Benchmarking the energy costs of large language model inference},
  author={Samsi, Siddharth and Zhao, Dan and McDonald, Joseph and Li, Baolin and Michaleas, Adam and Jones, Michael and Bergeron, William and Kepner, Jeremy and Tiwari, Devesh and Gadepally, Vijay},
  booktitle={IEEE High Performance Extreme Computing Conference (HPEC)},
  pages={1--9},
  year={2023},
  organization={IEEE}
}

@inproceedings{lazuka2024llm,
  title={Llm-pilot: Characterize and optimize performance of your llm inference services},
  author={Lazuka, Malgorzata and Anghel, Andreea and Parnell, Thomas},
  booktitle={SC24: International Conference for High Performance Computing, Networking, Storage and Analysis},
  pages={1--18},
  year={2024},
  organization={IEEE}
}

@inproceedings{martinez2025impact,
  title={The impact of hyperparameters on large language model inference performance: An evaluation of vllm and huggingface pipelines},
  author={Martinez, Matias},
  booktitle={Proceedings of the 33rd ACM International Conference on the Foundations of Software Engineering},
  pages={1672--1678},
  year={2025}
}

@article{xia2024top,
  title={Top leaderboard ranking= top coding proficiency, always? evoeval: Evolving coding benchmarks via llm},
  author={Xia, Chunqiu Steven and Deng, Yinlin and Zhang, Lingming},
  journal={arXiv preprint arXiv:2403.19114},
  year={2024}
}

@article{zhao2024wildchat,
  title={Wildchat: 1m chatgpt interaction logs in the wild},
  author={Zhao, Wenting and Ren, Xiang and Hessel, Jack and Cardie, Claire and Choi, Yejin and Deng, Yuntian},
  journal={preprint arXiv:2405.01470},
  year={2024}
}

@article{coignion2024green,
  title={Green My LLM: Studying the key factors affecting the energy consumption of code assistants},
  author={Coignion, Tristan and Quinton, Cl{\'e}ment and Rouvoy, Romain},
  journal={arXiv preprint arXiv:2411.11892},
  year={2024}
}

@inproceedings{bai2025longbench,
  title={Longbench v2: Towards deeper understanding and reasoning on realistic long-context multitasks},
  author={Bai, Yushi and Tu, Shangqing and Zhang, Jiajie and Peng, Hao and Wang, Xiaozhi and Lv, Xin and Cao, Shulin and Xu, Jiazheng and Hou, Lei and Dong, Yuxiao and others},
  booktitle={Proceedings of the 63rd Annual Meeting of the Association for Computational Linguistics (Volume 1: Long Papers)},
  pages={3639--3664},
  year={2025}
}

@article{kwiatkowski2019natural,
  title={Natural questions: a benchmark for question answering research},
  author={Kwiatkowski, Tom and Palomaki, Jennimaria and Redfield, Olivia and Collins, Michael and Parikh, Ankur and Alberti, Chris and Epstein, Danielle and Polosukhin, Illia and Devlin, Jacob and Lee, Kenton and others},
  journal={Transactions of the Association for Computational Linguistics},
  volume={7},
  pages={453--466},
  year={2019},
  publisher={MIT Press One Rogers Street, Cambridge, MA 02142-1209, USA journals-info~…}
}

@misc{lighteval,
  author = {Habib, Nathan and Fourrier, Clémentine and Kydlíček, Hynek and Wolf, Thomas and Tunstall, Lewis},
  title = {LightEval: A lightweight framework for LLM evaluation},
  year = {2023},
  version = {0.11.0},
  url = {https://github.com/huggingface/lighteval}
}

@misc{vllm_docs,
  title = {vLLM -- Configuration options},
  year = {2026},
  version = {0.17.1},
  url = {https://docs.vllm.ai/en/latest/configuration}
}

@misc{vllmOptimizationDocs,
  title = {vLLM -- Optimization and Tuning},
  year = {2026},
  version = {0.17.1},
  url = {https://docs.vllm.ai/en/latest/configuration/optimization.html}
}

@article{chen2021evaluating,
  title={Evaluating large language models trained on code},
  author={Chen, Mark and Tworek, Jerry and Jun, Heewoo and Yuan, Qiming and Pinto, Henrique Ponde De Oliveira and Kaplan, Jared and Edwards, Harri and Burda, Yuri and Joseph, Nicholas and Brockman, Greg and others},
  journal={arXiv preprint arXiv:2107.03374},
  year={2021}
}

@article{dao2022flashattention,
  title={Flashattention: Fast and memory-efficient exact attention with io-awareness},
  author={Dao, Tri and Fu, Dan and Ermon, Stefano and Rudra, Atri and R{\'e}, Christopher},
  journal={Advances in neural information processing systems},
  volume={35},
  pages={16344--16359},
  year={2022}
}

@article{dao2023flashattention,
  title={Flashattention-2: Faster attention with better parallelism and work partitioning},
  author={Dao, Tri},
  journal={arXiv preprint arXiv:2307.08691},
  year={2023}
}

@article{ye2025flashinfer,
  title={Flashinfer: Efficient and customizable attention engine for llm inference serving},
  author={Ye, Zihao and Chen, Lequn and Lai, Ruihang and Lin, Wuwei and Zhang, Yineng and Wang, Stephanie and Chen, Tianqi and Kasikci, Baris and Grover, Vinod and Krishnamurthy, Arvind and others},
  journal={arXiv preprint arXiv:2501.01005},
  year={2025}
}

@article{fu2025llmco2,
  title={Llmco2: Advancing accurate carbon footprint prediction for llm inferences},
  author={Fu, Zhenxiao and Chen, Fan and Zhou, Shan and Li, Haitong and Jiang, Lei},
  journal={ACM SIGENERGY Energy Informatics Review},
  volume={5},
  number={2},
  pages={63--68},
  year={2025},
  publisher={ACM New York, NY, USA}
}

@incollection{ollama,
  title={Using ollama},
  author={Marcondes, Francisco S and Gala, Adelino and Magalh{\~a}es, Renata and Perez de Britto, Fernando and Dur{\~a}es, Dalila and Novais, Paulo},
  booktitle={Natural Language Analytics with Generative Large-Language Models: A Practical Approach with Ollama and Open-Source LLMs},
  pages={23--35},
  year={2025},
  publisher={Springer}
}

@inproceedings{llms_landscape,
  title={Llm inference serving: Survey of recent advances and opportunities},
  author={Li, Baolin and Jiang, Yankai and Gadepally, Vijay and Tiwari, Devesh},
  booktitle={2024 IEEE High Performance Extreme Computing Conference (HPEC)},
  pages={1--8},
  year={2024},
  organization={IEEE}
}

@INPROCEEDINGS{pareto,
  author={Ngatchou, P. and Zarei, A. and El-Sharkawi, A.},
  booktitle={Proceedings of the 13th International Conference on, Intelligent Systems Application to Power Systems}, 
  title={Pareto Multi Objective Optimization}, 
  year={2005},
  volume={},
  number={},
  pages={84-91},
  doi={10.1109/ISAP.2005.1599245}}

@article{vargha2000critique,
  title={A critique and improvement of the CL common language effect size statistics of McGraw and Wong},
  author={Vargha, Andr{\'a}s and Delaney, Harold D},
  journal={Journal of Educational and Behavioral Statistics},
  volume={25},
  number={2},
  pages={101--132},
  year={2000},
  publisher={Sage Publications Sage CA: Los Angeles, CA}
}

@article{cliff1993dominance,
  title={Dominance statistics: Ordinal analyses to answer ordinal questions.},
  author={Cliff, Norman},
  journal={Psychological bulletin},
  volume={114},
  number={3},
  pages={494},
  year={1993},
  publisher={American Psychological Association}
}

@article{hu2025taming,
  title={Taming the long-tail: Efficient reasoning rl training with adaptive drafter},
  author={Hu, Qinghao and Yang, Shang and Guo, Junxian and Yao, Xiaozhe and Lin, Yujun and Gu, Yuxian and Cai, Han and Gan, Chuang and Klimovic, Ana and Han, Song},
  journal={preprint arXiv:2511.16665},
  year={2025}
}

@article{ait2025suitability,
  title={On the suitability of hugging face hub for empirical studies},
  author={Ait, Adem and C{\'a}novas Izquierdo, Javier Luis and Cabot, Jordi},
  journal={Empirical Software Engineering},
  volume={30},
  number={2},
  pages={57},
  year={2025},
  publisher={Springer}
}

@misc{huggingfaceTextGeneration,
	author = {},
	title = {{T}ext {G}eneration {M}odels – {H}ugging {F}ace --- huggingface.co},
	howpublished = {\url{https://huggingface.co/models?pipeline\_tag=text-generation\&sort=downloads}},
	year = {2025},
}

@misc{llamacpp,
	title = {ggml-org/llama.cpp},
	copyright = {MIT},
	url = {https://github.com/ggml-org/llama.cpp},
	urldate = {2026-03-06},
	publisher = {ggml},
	month = mar,
	year = {2026}
}

@article{wolf2019huggingface,
  title={Huggingface's transformers: State-of-the-art natural language processing},
  author={Wolf, Thomas and Debut, Lysandre and Sanh, Victor and Chaumond, Julien and Delangue, Clement and Moi, Anthony and Cistac, Pierric and Rault, Tim and Louf, R{\'e}mi and Funtowicz, Morgan and others},
  journal={arXiv preprint arXiv:1910.03771},
  year={2019}
}

@article{su2025seesaw,
  title={Seesaw: High-throughput llm inference via model re-sharding},
  author={Su, Qidong and Zhao, Wei and Li, Xin and Andoorveedu, Muralidhar and Jiang, Chenhao and Zhu, Zhanda and Song, Kevin and Giannoula, Christina and Pekhimenko, Gennady},
  journal={preprint arXiv:2503.06433},
  year={2025}
}

@inproceedings{wobbrock2011aligned,
  title={The aligned rank transform for nonparametric factorial analyses using only anova procedures},
  author={Wobbrock, Jacob O and Findlater, Leah and Gergle, Darren and Higgins, James J},
  booktitle={Proceedings of the SIGCHI conference on human factors in computing systems},
  pages={143--146},
  year={2011}
}

@article{wu2022sustainable,
  title={Sustainable ai: Environmental implications, challenges and opportunities},
  author={Wu, Carole-Jean and Raghavendra, Ramya and Gupta, Udit and Acun, Bilge and Ardalani, Newsha and Maeng, Kiwan and Chang, Gloria and Aga, Fiona and Huang, Jinshi and Bai, Charles and others},
  journal={Proceedings of machine learning and systems},
  volume={4},
  pages={795--813},
  year={2022}
}

@article{delavande2026understanding,
  title={Understanding Efficiency: Quantization, Batching, and Serving Strategies in LLM Energy Use},
  author={Delavande, Julien and Pierrard, Regis and Luccioni, Sasha},
  journal={arXiv preprint arXiv:2601.22362},
  year={2026}
}

@inproceedings{zine2026pimp,
  TITLE = {{Pimp My LLM: Leveraging Variability Modeling to Tune Inference Hyperparameters}},
  AUTHOR = {Zine, Nada and Quinton, Cl{\'e}ment and Rouvoy, Romain},
  URL = {https://hal.science/hal-05567430},
  BOOKTITLE = {{EASE'26 - 30th International Conference on Evaluation and Assessment in Software Engineering}},
  ADDRESS = {Glasgow, United Kingdom},
  YEAR = {2026},
  MONTH = Jun,
  HAL_ID = {hal-05567430},
  HAL_VERSION = {v1},
}

@article{shi2402thorough,
  title={A thorough examination of decoding methods in the era of llms, 2024},
  author={Shi, Chufan and Yang, Haoran and Cai, Deng and Zhang, Zhisong and Wang, Yifan and Yang, Yujiu and Lam, Wai},
  journal={URL https://arxiv. org/abs/2402.06925}
}

@inproceedings{arias2025decoding,
  title={Decoding decoded: Understanding hyperparameter effects in open-ended text generation},
  author={Arias, Esteban Garces and Li, Meimingwei and Heumann, Christian and A{\ss}enmacher, Matthias},
  booktitle={Proceedings of the 31st International Conference on Computational Linguistics},
  pages={9992--10020},
  year={2025}
}

@article{nik2025impact,
  title={Impact of decoding strategies on GPU energy usage in large language model text generation},
  author={Nik, Alireza and Riegler, Michael A and Halvorsen, P{\aa}l},
  journal={Scientific Reports},
  year={2025},
  publisher={Nature Publishing Group}
}

@article{holm1979simple,
  title={A simple sequentially rejective multiple test procedure},
  author={Holm, Sture},
  journal={Scandinavian journal of statistics},
  pages={65--70},
  year={1979},
  publisher={JSTOR}
}

@inproceedings{kolovskaSmallPromptsBig2025,
  title={Small Prompts, Big Energy and CO 2 Impact: Benchmarking Ollama LLMs on CPU and GPU},
  author={Kolovska, Ana and Gusev, Marjan and Mileski, Dimitar},
  booktitle={2025 33rd Telecommunications Forum (TELFOR)},
  pages={1--4},
  year={2025},
  organization={IEEE}
}

@article{niuEnergyEfficientExhaustive2025,
  title = {Energy {{Efficient}} or {{Exhaustive}}? {{Benchmarking Power Consumption}} of {{LLM Inference Engines}}},
  shorttitle = {Energy {{Efficient}} or {{Exhaustive}}?},
  author = {Niu, Chenxu and Zhang, Wei and Zhao, Yongjian and Chen, Yong},
  year = 2025,
  journal = {SIGENERGY Energy Inform. Rev.},
  volume = {5},
  number = {2},
  pages = {56--62},
  doi = {10.1145/3757892.3757900},
  urldate = {2025-09-17}
}

@misc{pronkBenchmarkingEnergyEfficiency2025,
  title = {Benchmarking {{Energy Efficiency}} of {{Large Language Models Using vLLM}}},
  author = {Pronk, K. and Zhao, Q.},
  year = 2025,
  month = sep,
  number = {arXiv:2509.08867},
  eprint = {2509.08867},
  primaryclass = {cs},
  publisher = {arXiv},
  doi = {10.48550/arXiv.2509.08867},
  urldate = {2025-09-19},
  archiveprefix = {arXiv}
}

@misc{jagannadharaoBeginnersGuidePower2024,
  title = {A {{Beginner}}'s {{Guide}} to {{Power}} and {{Energy Measurement}} and {{Estimation}} for {{Computing}} and {{Machine Learning}}},
  author = {Jagannadharao, Akshaya and Beckage, Nicole and Biswas, Sovan and Egan, Hilary and Gafur, Jamil and Metsch, Thijs and Nafus, Dawn and Raffa, Giuseppe and Tripp, Charles},
  year = 2024,
  month = dec,
  number = {arXiv:2412.17830},
  eprint = {2412.17830},
  primaryclass = {eess},
  publisher = {arXiv},
  doi = {10.48550/arXiv.2412.17830},
  urldate = {2025-03-20},
  archiveprefix = {arXiv}
}

@misc{holtzmanCuriousCaseNeural2020,
  title = {The {{Curious Case}} of {{Neural Text Degeneration}}},
  author = {Holtzman, Ari and Buys, Jan and Du, Li and Forbes, Maxwell and Choi, Yejin},
  year = 2020,
  month = feb,
  number = {arXiv:1904.09751},
  eprint = {1904.09751},
  primaryclass = {cs},
  publisher = {arXiv},
  urldate = {2023-02-23},
  archiveprefix = {arXiv}
}

@misc{baltesGuidelinesEmpiricalStudies2025,
  title = {Guidelines for {{Empirical Studies}} in {{Software Engineering}} Involving {{Large Language Models}}},
  author = {Baltes, Sebastian and Angermeir, Florian and Arora, Chetan and Bar{\'o}n, Marvin Mu{\~n}oz and Chen, Chunyang and B{\"o}hme, Lukas and Calefato, Fabio and Ernst, Neil and Falessi, Davide and Fitzgerald, Brian and Fucci, Davide and Kalinowski, Marcos and Lambiase, Stefano and Russo, Daniel and Lungu, Mircea and Prechelt, Lutz and Ralph, Paul and van Tonder, Rijnard and Treude, Christoph and Wagner, Stefan},
  year = 2025,
  month = sep,
  number = {arXiv:2508.15503},
  eprint = {2508.15503},
  primaryclass = {cs},
  publisher = {arXiv},
  doi = {10.48550/arXiv.2508.15503},
  urldate = {2025-09-17},
  archiveprefix = {arXiv}
}

@article{solovyeva2026towards,
  title={Towards Green AI: Decoding the Energy of LLM Inference in Software Development},
  author={Solovyeva, Lola and Castor, Fernando},
  journal={preprint arXiv:2602.05712},
  year={2026}
}

\end{document}